\documentclass[%
reprint,
superscriptaddress, 
amsmath,amssymb,
aps,
prb,
prstab,
floatfix,
longbibliography
]{revtex4-2}

\usepackage[caption=false]{subfig}
\usepackage{multirow}
\usepackage{graphicx}
\usepackage{dcolumn}
\usepackage{bm}
\usepackage{threeparttable}
\usepackage{color}
\usepackage[dvipsnames]{xcolor}
\usepackage{chemformula}
\usepackage{xspace}
\usepackage{orcidlink}
\usepackage{mathtools} 
\usepackage{extarrows} 

\usepackage{tabularx}
\usepackage{multirow}
\usepackage{graphicx}
\usepackage{dcolumn}
\usepackage{bm}
\usepackage{threeparttable}
\usepackage{booktabs}
\usepackage{color}
\usepackage{booktabs}
\usepackage{xspace}
\usepackage{chemformula}
\usepackage{xcolor}
\usepackage{scrextend}
\usepackage{hyperref}
\usepackage{orcidlink}

\hypersetup{colorlinks=true,linkcolor=blue,filecolor=blue,urlcolor=blue,citecolor=blue,pdftitle = {Lattice dynamics and mixing of polar phonons in the rare-earth orthoferrite TbFeO$_{3}$}, pdfauthor = {R.~M.~Dubrovin}}

\definecolor{myblue}{RGB}{0,0,0}
\definecolor{ER}{RGB}{0,0,0}
\definecolor{newtext}{RGB}{0,0,0}

\newcommand{\TFO}{$\mathrm{TbFeO}_{3}$}

\newcommand{\RFO}{$\mathrm{RFeO}_{3}$}

\bibpunct{[}{]}{;}{n}{}{}

\begin{document}

\title{Lattice dynamics and mixing of polar phonons in the rare-earth orthoferrite TbFeO$_{3}$}

\author{R.~M.~Dubrovin\,\orcidlink{0000-0002-7235-7805}}
\email{dubrovin@mail.ioffe.ru}
\affiliation{Ioffe Institute, Russian Academy of Sciences, 194021 St.\,Petersburg, Russia}
\author{E.~M.~Roginskii\,\orcidlink{0000-0002-5627-5877}}
\affiliation{Ioffe Institute, Russian Academy of Sciences, 194021 St.\,Petersburg, Russia}
\author{V.~A.~Chernyshev\,\orcidlink{0000-0002-3106-3069}}
\affiliation{Department of Basic and Applied Physics, Ural Federal University, 620002 Yekaterinburg, Russia}
\author{N.~N.~Novikova\,\orcidlink{0000-0003-2428-6114}}
\affiliation{Institute of Spectroscopy, Russian Academy of Sciences, 108840 Moscow, Troitsk, Russia}
\author{M.~A.~Elistratova\,\orcidlink{0000-0002-1573-1151}}
\affiliation{Ioffe Institute, Russian Academy of Sciences, 194021 St.\,Petersburg, Russia}
\author{I.~A.~Eliseyev\,\orcidlink{0000-0001-9980-6191}}
\affiliation{Ioffe Institute, Russian Academy of Sciences, 194021 St.\,Petersburg, Russia}
\author{A.~N.~Smirnov\,\orcidlink{0000-0001-9709-5138}}
\affiliation{Ioffe Institute, Russian Academy of Sciences, 194021 St.\,Petersburg, Russia}
\author{A.~I.~Brulev\,\orcidlink{0009-0004-5339-8486}}
\affiliation{Ioffe Institute, Russian Academy of Sciences, 194021 St.\,Petersburg, Russia}
\affiliation{University of Nizhny Novgorod, 603022 Nizhny Novgorod, Russia}
\author{K.~N.~Boldyrev\,\orcidlink{0000-0002-3784-7294}}
\affiliation{Institute of Spectroscopy, Russian Academy of Sciences, 108840 Moscow, Troitsk, Russia}
\author{V.~{Yu}.~Davydov\,\orcidlink{0000-0002-5255-9530}}
\affiliation{Ioffe Institute, Russian Academy of Sciences, 194021 St.\,Petersburg, Russia}
\author{R.~V.~Mikhaylovskiy\,\orcidlink{0000-0003-3780-0872}}
\affiliation{Department of Physics, Lancaster University, Bailrigg, Lancaster LA1 4YW, United Kingdom}
\author{A.~M.~Kalashnikova\,\orcidlink{0000-0001-5635-6186}}
\affiliation{Ioffe Institute, Russian Academy of Sciences, 194021 St.\,Petersburg, Russia}
\author{R.~V.~Pisarev\,\orcidlink{0000-0002-2008-9335}}
\affiliation{Ioffe Institute, Russian Academy of Sciences, 194021 St.\,Petersburg, Russia}

\date{\today}

\begin{abstract}
Rare-earth orthoferrites are a promising platform for antiferromagnetic spintronics with a rich variety of terahertz spin and lattice dynamics phenomena.
For instance, it has been experimentally demonstrated that the light-driven optical phonons can coherently manipulate macroscopic magnetic states via nonlinear magnetophononic effects.
Here using TbFeO$_{3}$ as an example, we reveal the origin of the mode mixing between the LO and TO phonons, which is important for understanding of nonlinear phononics. 
We performed a comprehensive study of the lattice dynamics of the TbFeO$_{3}$ single crystal by polarized infrared and Raman scattering spectroscopic techniques, and experimentally obtained and carefully analyzed the spectra of anisotropic complex dielectric functions in the far-infrared spectral range. 
This allowed us to reliably identify the symmetries and parameters of most infrared- and Raman-active phonons. 
Next, the experimental studies were supplemented by the lattice dynamics calculations which allowed us to propose the normal mode assignments.
We reveal that the relation between LO and TO polar phonons is complex and does not strictly follow the ``LO-TO rule'' due to the strong mode mixing. 
We further analyze how displacements of different ions contribute to phonon modes and reveal that magnetic Fe ions are not involved in Raman-active phonons, thus shedding light on a lack of spin phonon coupling for such phonons.
The obtained results establish a solid basis for further in-depth experimental research in the field of nonlinear phononics and magnetophononics in rare-earth orthoferrites.
\end{abstract}

\maketitle

\section{Introduction}
\label{sec:intro}
Rare-earth orthoferrites \RFO, where $\mathrm{R}$ stands for a rare-earth cation, are a universe for researchers in the area of spin physics because of the many exciting magnetic~\cite{leenders2024canted,kimel2023universal,li2023terahertz,afanasiev2021ultrafast,li2018observation,grishunin2018terahertz,nova2017effective,artyukhin2012solitonic,johnson2022spiers,kimel2009inertia,kimel2005ultrafast,kimel2004laser,moskvin2021dzyaloshinskii,moskvin2022simple}, magnetoelectric~\cite{yamaguchi1973magnetic,zvezdin2008magnetoelectric,tokunaga2009composite,stanislavchuk2017far,alizera2022origin,ivanov2023observation}, multiferroic~\cite{hassanpour2022magnetoelectric}, and other properties observed in them.
The orthoferrites \RFO{} have been known for over 60 years and in many ways have already become well characterized model materials, but nevertheless their potential has not been fully realized and they are still a universal playground for modern magnetism~\cite{smejkal2022emerging}. 
The presence of the $3d$ $\mathrm{Fe^{3+}}$ and $4f$ $\mathrm{R^{3+}}$ magnetic cations in different sublattices leads to competition between $\mathrm{Fe}-\mathrm{Fe}$, $\mathrm{R}-\mathrm{Fe}$, and $\mathrm{R}-\mathrm{R}$ exchange interactions and, in turn, to a complex magnetic phase diagram with a variety of spin-reorientation transitions~\cite{belov1976spin}.
Thus, the control of macroscopic magnetic states in rare-earth orthoferrites creates a rich platform for application in high speed data storage devices~\cite{leenders2024canted,zhang2024upconversion,zhang2024coupling,huang2024extreme,li2023terahertz,han2023coherent,kurihara2023observation,das2022anisotropic,kimel2020fundamentals}.

It is known that lattice dynamics is responsible for important physical properties of crystals such as thermodynamical characteristics, superconductivity, and phase transitions~\cite{dove1993introduction}.
Moreover, nowadays the resonant driving of phonons in crystals is a unique route for coherent manipulation of the lattice and its associated functional properties at high rates, which is not available in equilibrium~\cite{disa2021engineering}.
Meanwhile, orthoferrites are archetypical magneto-phononic materials in which it has been experimentally shown for the first time that resonant excitation of polar phonons using intense infrared light provides a unique opportunity for coherent control of macroscopic magnetic states~\cite{nova2017effective,juraschek2017ultrafast,afanasiev2021ultrafast}.
The lattice dynamics of orthoferrites \RFO{} in the center of the Brillouin zone has been studied in depth by numerical simulations~\cite{gupta2002lattice,singh2008polarized,juraschek2017ultrafast,wang2019first,ahmed2021structural} and Raman spectroscopy~\cite{koshizuka1980inelastic,venugopalan1985magnetic,singh2008polarized,mihalik2015heat,weber2016raman,panchwanee2017study,coutinho2017structural,panchwanee2019temperature,saha2020structure,ye2020temperature,ponosov2020lattice,khan2021magneto,weber2022emerging,mali2022spin,eymeoud2023raman}.
In contrast, the infrared-active phonons in orthoferrites \RFO{} have been studied in most cases in non-single-crystal samples~\cite{rao1970infrared,mathe2004synthesis,jamil2018optical,haye2018experimental,song2018effects,suthar2020synthesis,massa2023low} and the studies carried out on single-crystal samples did not concern polarizations along the main crystallographic axes~\cite{tajima1987infrared,laforge2013electron,nova2017effective}, which makes it almost impossible to establish the symmetry of the studied polar phonons using the selection rules for polarization of radiation.
To our knowledge, there is only one recent paper with results of the polarization-resolved measurements on the orthoferrite single crystal~\cite{komandin2023electric}.

\begin{figure*}
\centering
\includegraphics[width=2\columnwidth]{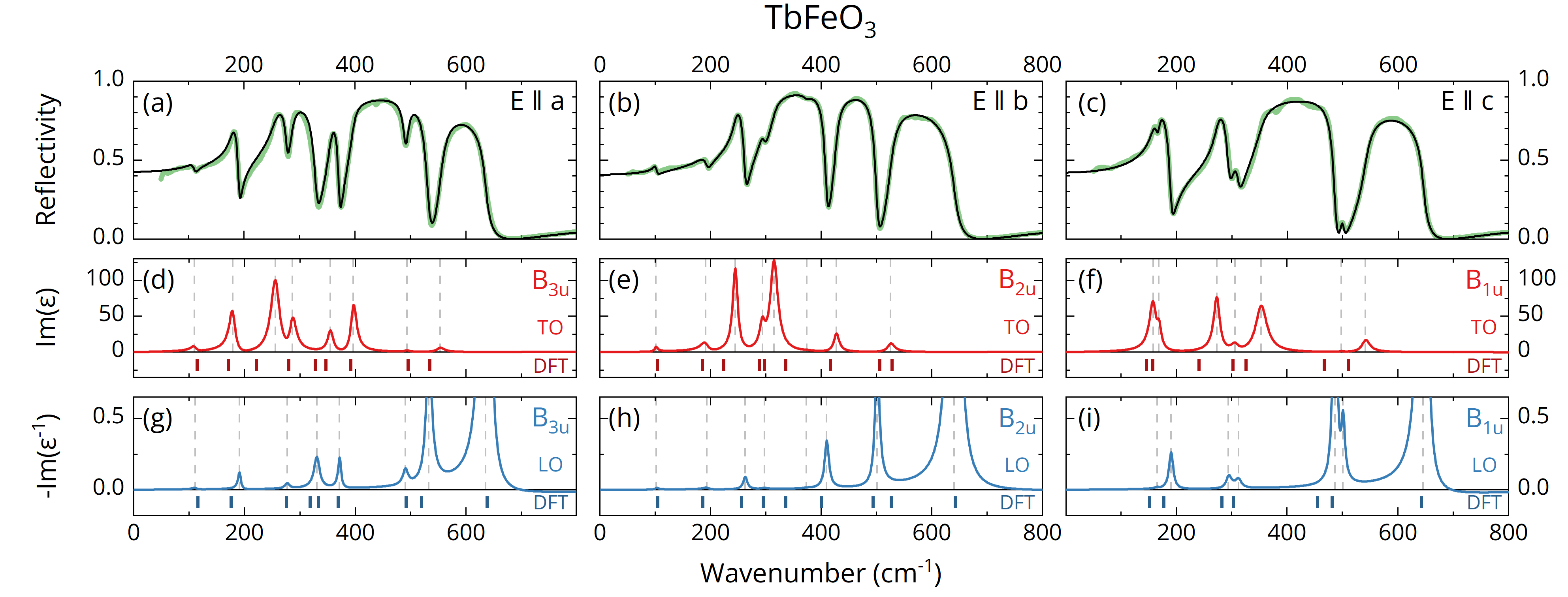}
\caption{\label{fig:reflectivity}
The polarized infrared reflectivity spectra at ambient conditions with the electric field of light $\textbf{E}$ polarized along the (a)~$a$, (b)~$b$, and (c)~$c$ axes of the orthoferrite $\mathrm{TbFeO}_{3}$.
The solid black lines are fits based on the generalized oscillator model of the complex dielectric permittivity $\varepsilon(\omega) = \varepsilon_{1}(\omega) - i\varepsilon_{2}(\omega)$ according to Eq.~(S1)  in SM~\cite{supp_mat}.
Spectra of the imaginary part of the complex dielectric permittivity $\Im[\varepsilon(\omega)]$ and the inverse complex dielectric permittivity $-\Im[\varepsilon^{-1}(\omega)]$ obtained from the fit of the reflectivity spectra corresponding to the TO and LO polar phonons with the symmetries (d), (g)~$B_{3u}$, (e), (h)~$B_{2u}$, and (f), (i)~$B_{1u}$, respectively.
Red and blue tick marks present the calculated TO and LO frequencies of the polar phonons, respectively.
}
\end{figure*}

In this paper, we present results of a systematic study of the lattice dynamics of orthoferrite \TFO{} high-quality single crystals employing complementary infrared reflectivity and Raman scattering polarized spectroscopic techniques supported by corresponding ab initio calculations.
The anisotropic complex dielectric function was accurately extracted from spectroscopic reflectivity measurements at infrared frequencies for the main crystallographic axes of the studied orthoferrite.
A rigorous examination of the obtained experimental spectra allowed us to successfully identify frequencies and symmetries of most of the infrared- and Raman-active phonons.
Moreover, the analysis of the calculated eigendisplacements allowed us to establish the couplings between LO and TO phonons which sets the grounds for further exploiting of phononics in this class of crystals. 
\textcolor{newtext}{Note that our research is focused on the room temperature because all rare-earth orthoferrites except $\mathrm{SmFeO}_{3}$ have the same $\Gamma_{4}$ magnetic configuration at ambient conditions~\cite{li2023terahertz} which is important for application in nonlinear phononics and therefore the conclusions of the paper can be extended to all these orthoferrites.}

This paper is organized as follows.
\textcolor{newtext}{The samples of \TFO{} and an outline of the experimental setups and computational details for studying the lattice dynamics are introduced in the Supplemental Material (SM)}~\cite{[{See Supplemental Material at }] [{ for details on the samples of TbFeO$_{3}$, experimental setups and computational details for studying the lattice dynamics and its analysis.
The Supplemental Material contains  Refs.~\cite{balbashov2019contemporary,balbashov1981apparatus,perdew1996generalized,kresse1996efficiency,kresse1996efficient,dudarev1998electron,monkhorst1976special,togo2015first,wang2010mixed,maschio2012infrared,maschio2013raman,dovesi2014crystal14,becke1993density,dolg1989energy,dolg1993combination,yang2005valence,peintinger2013consistent,kuzmenko2005kramers,schubert2004infrared,gervais1974anharmonicity,lyddane1941polar,born2013principles,martin2001melting,kroumova2003bilbao,loudon2001raman,damen1966raman}}] supp_mat}.
\nocite{balbashov2019contemporary,balbashov1981apparatus,perdew1996generalized,kresse1996efficiency,kresse1996efficient,dudarev1998electron,monkhorst1976special,togo2015first,wang2010mixed,maschio2012infrared,maschio2013raman,dovesi2014crystal14,becke1993density,dolg1989energy,dolg1993combination,yang2005valence,peintinger2013consistent,kuzmenko2005kramers,schubert2004infrared,gervais1974anharmonicity,lyddane1941polar,born2013principles,martin2001melting,kroumova2003bilbao,loudon2001raman,damen1966raman}
Section~\ref{sec:results} presents and discusses the experimental data on infrared and Raman spectroscopy, supported by first-principles calculations, which allow us to reveal the strong LO-TO mixing and follow the phonon genesis.
Concluding remarks are given in Sec.~\ref{sec:conclusion}.

\section{Results and discussion}
\label{sec:results}

\begin{table*}
\caption{\label{tab:irrep} Character table of irreducible representations of the $D_{2h}(mmm)$ point group in the $Pbnm$ coordinate axes orientation.}
\begin{ruledtabular}
\begin{tabular}{cccccccccc}
$D_{2h}$ & $E$ & $C_{2}(x)$ & $C_{2}(y)$ & $C_{2}(z)$ & $i$ & $\sigma(xy)$ & $\sigma(yz)$ & $\sigma(xz)$ & functions \\ \hline
 $A_{g}$ & $+1$ & $+1$ & $+1$ & $+1$ & $+1$ & $+1$ & $+1$ & $+1$ & $x^{2}$, $y^{2}$, $z^{2}$ \\
$B_{1g}$ & $+1$ & $-1$ & $-1$ & $+1$ & $+1$ & $+1$ & $-1$ & $-1$ & $R_{z}$, $xy$ \\
$B_{2g}$ & $+1$ & $-1$ & $+1$ & $-1$ & $+1$ & $-1$ & $-1$ & $+1$ & $R_{y}$, $xz$ \\
$B_{3g}$ & $+1$ & $+1$ & $-1$ & $-1$ & $+1$ & $-1$ & $+1$ & $-1$ & $R_{x}$, $yz$ \\
 $A_{u}$ & $+1$ & $+1$ & $+1$ & $+1$ & $-1$ & $-1$ & $-1$ & $-1$ & $xyz$ \\
$B_{1u}$ & $+1$ & $-1$ & $-1$ & $+1$ & $-1$ & $-1$ & $+1$ & $+1$ & $z$, $x^{2}z$, $y^{2}z$, $z^{3}$ \\
$B_{2u}$ & $+1$ & $-1$ & $+1$ & $-1$ & $-1$ & $+1$ & $+1$ & $-1$ & $y$, $yx^{2}$, $y^{3}$, $yz^{2}$ \\
$B_{3u}$ & $+1$ & $+1$ & $-1$ & $-1$ & $-1$ & $+1$ & $-1$ & $+1$ & $x$, $x^{3}$, $xz^{2}$, $xy^{2}$ \\
\end{tabular}
\end{ruledtabular}
\end{table*}

Rare-earth orthoferrite \TFO{} has the orthorhombic crystal structure with the space group $Pbnm$ [\#62, $D^{16}_{2h}$, $Pnma$ with nonconventional coordinate axes orientation, $Pnma(a,b,c) \Leftrightarrow Pbnm(b,c,a)$] and four formula units per unit cell $Z=4$~\cite{eibschutz1965lattice,marezio1970crystal}. 
The lattice parameters \textcolor{newtext}{measured by x-ray diffraction} at room temperature \textcolor{newtext}{in the $Pbnm$ coordinate axes orientation} are $a=5.33$\,\AA, $b=5.6$\,\AA, and $c=7.65$\,\AA{} which are close to the literature data~\cite{stanislavchuk2017far}. 
The unit cell contains 20 ions occupying the Wyckoff positions $4c$ for \ch{Tb^{3+}}, $4b$ for \ch{Fe^{3+}}, $4c$ and $8d$ for \ch{O^{2-}}.
The orthorhombic structure of \TFO{} originates from the distortion of the ideal cubic perovskite structure with space group $Pm\overline{3}m$ caused by the ionic size mismatch~\cite{glazer1972classification}.

The group-theoretical analysis of $Pbnm$ orthoferrites \RFO{} predicts 60 phonons at the center of the Brillouin zone~\cite{kroumova2003bilbao}:
\begin{equation}
\label{eq:group_irrep_total_orth}
\begin{split} 
\Gamma_{\mathrm{total}} = \underbrace{B_{1u} \oplus B_{2u} \oplus B_{3u}}_{\Gamma_{\mathrm{acoustic}}} \oplus \underbrace{7 A_{g} \oplus 7 B_{1g} \oplus 5 B_{2g} \oplus 5 B_{3g}}_{\Gamma_{\mathrm{Raman}}} \oplus \\
\oplus \underbrace{7 B_{1u} \oplus 9 B_{2u} \oplus 9 B_{3u}}_{\Gamma_{\mathrm{IR}}} \oplus \underbrace{8A_{u}}_{\Gamma_{\mathrm{silent}}},
\end{split}
\end{equation}
among which there are 3 acoustic, 24 Raman-active, 25 infrared-active (polar), and 8 silent nondegenerate modes.
We note that silent modes are inactive neither in Raman nor in infrared spectra but can be observed in hyper-Raman experiments.
Table~\ref{tab:irrep} lists the characters for all modes.

\textcolor{newtext}{A comprehensive description of \TFO{} single crystal samples and an outline of the experimental setups and computational details for studying the lattice dynamics and features of their analysis are given in SM~\cite{supp_mat}.}

\subsection{Infrared spectroscopy}
\label{subsec:infra}

The reflectivity spectra of the orthoferrite \TFO{} measured at ambient conditions for the polarization of the electric field of light $\textbf{E}$ parallel to the $a$, $b$ and $c$ axes are shown by the green lines in Figs.~\ref{fig:reflectivity}(a)--\ref{fig:reflectivity}(c). 
According to the group-theoretical analysis~\cite{kroumova2003bilbao}, polar phonons with $B_{3u}$, $B_{2u}$, and $B_{1u}$ symmetries are active for electric field polarizations along the $a$, $b$, and $c$ axes, respectively. 
The reflection bands observed in the spectra allow us to readily identify 7 out of 7 $B_{1u}$, 8 out of 9 $B_{2u}$, and 8 out of 9 $B_{3u}$ polar phonons symmetry-allowed for corresponding polarizations.

\begin{table}
\caption{\label{tab:IR_phonons} Experimental frequencies $\omega$ (cm$^{-1}$), dampings $\gamma$ (cm$^{-1}$), and dielectric strengths $\Delta\varepsilon$ of the TO and LO polar phonons in \TFO{} at room temperature in comparison with the results of DFT calculations presented in parentheses.}
\begin{ruledtabular}
\begin{tabular}{ccccccccc}
Sym. & \multicolumn{2}{c}{$\omega_{\mathrm{TO}}$} & $\gamma_{\mathrm{TO}}$ & \multicolumn{2}{c}{$\omega_{\mathrm{LO}}$} & $\gamma_{\mathrm{LO}}$ & \multicolumn{2}{c}{$\Delta\varepsilon$}\\ \hline
\multirow{9}{*}{$B_{3u}$}  & 109.7 & (112.4) & 12.5 & 111   & (112.8) &  11  & 0.64 & (0.31)  \\
                           & 179.1 & (166.7) & 13.6 & 191.2 & (175.6) &  6.4 & 4.17 & (4.83)  \\ 
                           & 256.2 & (230.3) & 19.2 & 277.5 & (266.9) &  9.3 & 7.51 & (12.83) \\ 
                           & 286.6 & (274.7) &  14  & 330.8 & (314.1) & 12.3 & 2.17 & (1.48)  \\ 
                           & ---   & (322.5) & ---  & ---   & (323)   & ---  & ---  & (0.02)  \\ 
                           & 355.6 & (341.4) & 12.5 & 372   & (361.1) & 6.54 & 1.03 & (1.26)  \\ 
                           & 396.8 & (390.7) & 12.3 & 490.8 & (490.9) & 11.3 & 2.05 & (2.41)  \\ 
                           & 493.4 & (495)   & 11.5 & 532.9 & (507.3) & 12.4 & 0.04 & (0.04)  \\ 
                           & 553.7 & (521.5) & 17.2 & 635.7 & (610.8) & 11.3 & 0.19 & (0.22)  \\ \hline
\multirow{9}{*}{$B_{2u}$}  & 101.3 & (101.3) &  7.4 & 102.3 & (101.8) &  8   & 0.46 & (0.28)  \\
                           & 191.2 & (182.3) & 17.3 & 193.8 & (182.4) & 14.1 & 0.91 & (0.005) \\
                           & 245   & (226.2) &  11  & 262.8 & (249.8) &  8.7 & 5.18 & (9.44)  \\
                           & 294   & (286.5) & 11.9 & 297.5 & (287.5) & 11.9 & 1.38 & (1.21)  \\
                           & 314.8 & (294.6) & 16.8 & 410.2 & (396.3) &  9.2 & 6.74 & (8.54)  \\
                           & 374   & (329.6) & 15   & 373.6 & (329.6) & 14.8 & 0.03 & (-0.0002) \\
                           & 427.5 & (416.8) & 11.6 & 501.3 & (480.4) & 9.28 & 0.69 & (0.96)  \\
                           & ---   & (491.8) & ---  & ---   & (513.3) & ---  & ---  & (0.25)  \\
                           & 526   & (514.4) & 16.1 & 640.9 & (614.6) & 14.6 & 0.37 & (0.03)  \\ \hline
\multirow{7}{*}{$B_{1u}$}  & 158.7 & (149.4) & 15.2 & 166   & (153.1) &  8.7 & 6.43 & (6.64)  \\
                           & 168.7 & (155.7) &  8.3 & 190.8 & (174.3) &   9  & 1.53 & (3.2)   \\
                           & 273.6 & (243.7) & 13.5 & 294.4 & (284.6) & 12.1 & 3.73 & (8.33)  \\
                           & 306.1 & (297.2) & 16.1 & 312.6 & (297.3) & 13.2 & 0.49 & (0.003) \\
                           & 353.4 & (332.2) & 24.6 & 486.5 & (442.4) &  8.1 & 4.52 & (5.23)  \\
                           & 498.2 & (459.3) &  7.7 & 501.3 & (472.8) &  7.8 & 0.02 & (0.22)  \\
                           &  542  & (491.4) & 18.1 & 645.6 & (600.3) & 10.6 & 0.56 & (0.51)  \\
\end{tabular}
\end{ruledtabular}
\end{table}

There is a fair agreement between experimental spectra (green lines) and fits obtained using Eqs.~(S1)
and~(S2) in SM~\cite{supp_mat}
(black lines) seen for all studied polarizations in Figs.~\ref{fig:reflectivity}(a)--\ref{fig:reflectivity}(c).
The spectra of the $\Im[\varepsilon(\omega)]$ and $\Im[\varepsilon^{-1}(\omega)]$ corresponding to the fits are shown by red and blue lines in Figs.~\ref{fig:reflectivity}(d)--\ref{fig:reflectivity}(f) and~\ref{fig:reflectivity}(g)--\ref{fig:reflectivity}(i), respectively.
The frequencies and dampings of polar phonons for \TFO{} derived from the fits of the reflectivity spectra are listed in Table~\ref{tab:IR_phonons}.
It is worth noting that there is the generalized Lowndes condition $\sum_{j} (\gamma_{j\textrm{LO}} - \gamma_{j\textrm{TO}}) > 0$ that must be satisfied to keep positive $\Im[\varepsilon(\omega)]$ for  insulator crystals~\cite{lowndes1970influence,schubert2000infrared,schubert2004infrared}.
As follows from Table~\ref{tab:IR_phonons}, this condition is somewhat violated in our case of the best fits, but nevertheless no significant value of $\Im[\varepsilon(\omega)] < 0$ is observed as can be seen in Figs.~\ref{fig:reflectivity}(d)--\ref{fig:reflectivity}(i).

\begin{table}
\caption{\label{tab:epsilon} Experimental lattice parameters $a$, $b$ and $c$ (\AA) \textcolor{newtext}{obtained by the x-ray diffraction}, values of the static $\varepsilon_{0}$ and high frequency $\varepsilon_{\infty}$ anisotropic dielectric permittivities in \TFO{} at room temperature in comparison with the results of DFT calculations.}
\begin{ruledtabular}
\begin{tabular}{ccc}
                           & Exp   & DFT \\ \hline
$a$                        & 5.33 & 5.49 \\
$b$                        & 5.6  & 5.81 \\
$c$                        & 7.65 & 7.92 \\ \hline
$\varepsilon^{a}_{0}$      & 21.7 & 29.2 \\
$\varepsilon^{b}_{0}$      & 20.5 & 30.1 \\
$\varepsilon^{c}_{0}$      & 22.6 & 26.4 \\ \hline
$\varepsilon^{a}_{\infty}$ & 4.81 & 5.83 \\
$\varepsilon^{b}_{\infty}$ & 4.82 & 5.97 \\
$\varepsilon^{c}_{\infty}$ & 4.79 & 5.70 \\
\end{tabular}
\end{ruledtabular}
\end{table}

The contribution from each $j$th polar phonon of a specific symmetry to the anisotropic static dielectric permittivity $\varepsilon_{0} = \varepsilon_{\infty} + \sum_{j}\Delta\varepsilon_{j}$ is determined by its dielectric strength~\cite{gervais1983long}
\begin{equation}
\label{eq:oscillator_strength_TOLO}
\Delta\varepsilon_{j}  =  \frac{\varepsilon_{\infty}}{{\omega^{2}_{j\textrm{TO}}}}\frac{\prod\limits_{k}{\omega^{2}_{k\textrm{LO}}}-{\omega^{2}_{j\textrm{TO}}}}{\prod\limits_{k\neq{}j}{\omega^{2}_{k\textrm{TO}}}-{\omega^{2}_{j\textrm{TO}}}}.
\end{equation}
The TO and LO frequencies of polar phonons from the fits were used to obtain the values of dielectric strengths $\Delta\varepsilon$ by using Eq.~\eqref{eq:oscillator_strength_TOLO} which are listed in Table~\ref{tab:IR_phonons}.
The values of the anisotropic static $\varepsilon_{0}$ and high frequency $\varepsilon_{\infty}$ dielectric permittivities obtained from the reflectivity fits using Eq.~(S1) in SM~\cite{supp_mat} 
are listed in Table~\ref{tab:epsilon}.
This value of static dielectric permittivity $\varepsilon_{0}$ for \TFO{} is in fair agreement with data reported in the literature for several orthoferrites~\cite{balbashov1995submillimeter,stanislavchuk2016magnon}.


It should be noted that the analysis of the experimental results using Eq.~(S1) in SM~\cite{supp_mat} 
does not allow one to associate a given LO frequency with a TO frequency of the polar phonon with a particular symmetry.
Moreover, Eq.~\eqref{eq:oscillator_strength_TOLO} gives the same result for any relation between TO and LO frequencies of the polar phonons. 
However, there is the so-called ``TO-LO rule'' stating that for the each main crystallographic axis, the sequence of polar phonons is such that a TO frequency is always followed exactly by the corresponding LO frequency with an ascending frequency $\omega_{\textrm{LO}} > \omega_{\textrm{TO}}$, and therefore the LO-TO splitting is positive~\cite{schubert2019phonon}. 
Thus, applying this rule, the LO frequency can be assigned to the TO frequency of the polar phonon with a specific symmetry, which gives reliable results for many crystals.    
Besides, often the LO and TO frequencies are grouped according to similarity in strength and width of the peaks in spectra of $\Im[\varepsilon(\omega)]$ and $\Im[\varepsilon^{-1}(\omega)]$~\cite{fredrickson2016theoretical}.
However, none of these empirical rules is canonical in general and the lattice dynamical calculations are an efficient way to give reliable identification of LO-TO phonon pairs based on solid physical arguments.

\subsection{Raman spectroscopy}
\label{subsec:raman}

\begin{table}
\caption{\label{tab:Raman_phonons} Frequencies (cm$^{-1}$) and full widths at half maximum (FWHM, cm$^{-1}$) of the Raman-active $A_{g}$, $B_{1g}$, $B_{2g}$, and $B_{3g}$ phonons for \TFO{} at ambient conditions in comparison with results of DFT calculations and the experimental data from Refs.~\cite{weber2016raman,venugopalan1985magnetic}.
}
\begin{ruledtabular}
\begin{tabular}{ccccccc}
                          & \multicolumn{2}{c}{Experiment} & \multicolumn{2}{c}{DFT} & \multicolumn{2}{c}{Experiment} \\ \cmidrule{2-3} \cmidrule{4-5} \cmidrule{6-7}
Sym.                      & Freq. & FWHM & Freq.\footnote{\textsc{VASP}} & Freq.\footnote{\textsc{CRYSTAL14}} & Freq.\footnote{Ref.~\cite{weber2016raman}.} & Freq.\footnote{Ref.~\cite{venugopalan1985magnetic}.} \\ \hline
\multirow{7}{*}{$A_{g}$}  & 111   & 3.3  & 108.1 & 111.8 & 112.5 & 109 \\
                          & 139.7 & 6.5  & 130.5 & 131.8 & 143.9 & 140 \\
                          & 257.9 & 14   & 253.2 & 253.1 & 261.9 & 273 \\
                          & 331.6 & 7.4  & 319.2 & 330   & 334.5 & 329 \\
                          & 407.8 & 9.9  & 388.1 & 402.4 & 410.9 & --- \\ 
                          & 415.4 & 25.3 & 404.5 & 408.8 & 420.1 & 406 \\
                          & 487.2 & 10.8 & 471.6 & 487.6 & 490.1 & 480 \\ \hline
\multirow{7}{*}{$B_{1g}$} & 109.8 & 5.6  & 106.5 & 109.6 & 107.7 & --- \\
                          & 159.1 & 6.8  & 155.6 & 157.8 & 160.1 & 139 \\
                          & 296.6 & 14   & 297.2 & 293.9 & 302.7 & --- \\
                          & 359   & 9.9  & 338.8 & 351.7 & ---   & 329 \\
                          & 483.6 & 9.4  & 464.6 & 483.1 & 485.6 & 479 \\
                          & ---   & ---  & 520.1 & 529.3 & 535.8 & --- \\
                          & 639.6 & ---  & 608.3 & 641.1 & ---   & --- \\ \hline
\multirow{5}{*}{$B_{2g}$} & 128.1 & 6.2  & 118.2 & 122.6 & ---   & --- \\ 
                          & 321.5 & 12.2 & 297.8 & 314.3 & ---   & --- \\
                          & 428.1 & 20.2 & 382.1 & 412.4 & 433.3 & 418 \\
                          & 466.8 & ---  & 446.9 & 463.7 & 468.8 & --- \\
                          & ---   & ---  & 629.8 & 667.8 & ---   & --- \\ \hline
\multirow{5}{*}{$B_{3g}$} & 149.1 & 5.2  & 132.8 & 140.9 & ---   & 159 \\ 
                          & 252.4 & 17.7 & 235.6 & 239.8 & 251.9 & 249 \\
                          & 356.2 & 10.8 & 339.8 & 347.8 & 359.2 & 354 \\
                          & 426   & 20.7 & 393.2 & 409.6 & 427.7 & 426 \\
                          & 629.3 & 69.1 & 581.8 & 622.8 & ---   & --- \\
\end{tabular}
\end{ruledtabular}
\end{table}

\begin{figure*}
\centering
\includegraphics[width=2\columnwidth]{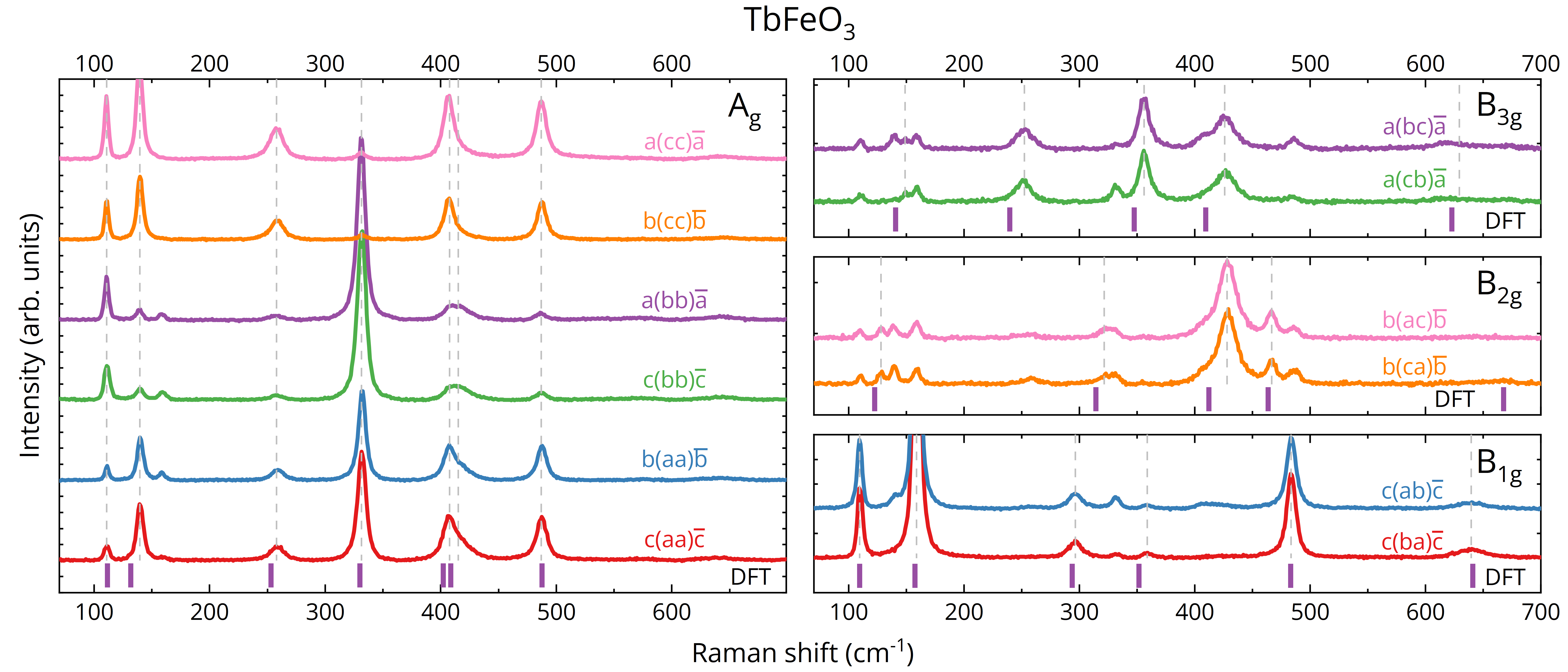}
\caption{\label{fig:raman}
Polarized Raman spectra of the $A_{g}$, $B_{3g}$, $B_{2g}$, and $B_{1g}$ phonons at ambient conditions for \TFO{}.
The polarization configurations are given in Porto's notation as described in the text.
Purple tick marks at the bottom of each plot present the calculated phonon frequencies.
}
\end{figure*}

The lattice dynamics at the Brillouin zone center in \TFO{} is studied further by analyzing of the optical phonons which are active in Raman scattering.
The experimental Raman spectra of \TFO{} in different polarizations measured at ambient conditions are presented in Fig.~\ref{fig:raman}.
The difference between the values of the diagonal elements of the Raman tensor~(S3) in SM~\cite{supp_mat} 
leads to unequal intensities of the fully symmetric $A_{g}$ phonons for various parallel polarizations.
The obtained Raman spectra have been carefully analyzed, and 7 out of 7 expected $A_{g}$ phonons, 6 out of 7 for $B_{1g}$, 4 out of 5 for $B_{2g}$, and 5 out of 5 for $B_{3g}$ modes were reliably identified as shown by dashed lines in Fig.~\ref{fig:raman}.
The frequencies, intensities, and full widths at half maximum (FWHMs) of the identified phonons were extracted by fitting of the obtained Raman spectra with a sum of Voigt profiles~\cite{wojdyr2010fityk} and listed in Table~\ref{tab:Raman_phonons}.
Small leaks of phonons in forbidden polarizations were observed due to the almost unavoidable depolarization effect in the optical elements and slight misalignment of the polarization of light with respect to the crystal axes. 
It is worth noting that the intensities of the $A_{g}$ modes are significantly higher than for the $B_{g}$ phonons, as shown in Fig.~\ref{fig:raman}.
The calculated frequencies of the Raman-active phonons in \TFO{} are in fair agreement with experimental values as listed in Table~\ref{tab:Raman_phonons}.
Besides, there is a fair agreement between the phonon frequencies obtained in our experiment and the ones from Refs.~\cite{weber2016raman,venugopalan1985magnetic}, as can be seen in Table~\ref{tab:Raman_phonons}.



To reveal the symmetry of the weak and overlapping lines we performed angle-resolved Raman measurements for both parallel ($\bm{e}_{i}\parallel{\bm{e}_{s}}$) and crossed ($\bm{e}_{i}\perp{\bm{e}_{s}}$) polarizations.
The experimental angular-dependent spectral intensity maps are shown in Fig.~S1 in SM~\cite{supp_mat}.
As expected, the phonon lines exhibit strong anisotropy of Raman scattering.
Note that the experimental intensity maps are in fair agreement with the results of the corresponding DFT calculations as can be seen in Figs.~S1 
and~S2 in SM~\cite{supp_mat}.



Further, the experimental angular dependences of phonon intensity were extracted (see colored open circles in Fig.~S3 in SM~\cite{supp_mat}). 
To verify the phonon symmetry, the obtained angular dependences were fitted using Eqs.~(S4)
and~(S3) 
as shown by colored solid lines in Fig.~S3 in SM~\cite{supp_mat}. 
It should be noted that 3 parallel and 3 crossed angular dependences are fitted at once using the same Raman tensor.
A satisfactory agreement between the experimental data and fit lines is observed as seen in Fig.~S3 in SM~\cite{supp_mat}. 
The $A_{g}$ modes have the highest intensity along the main crystal axes in the parallel configuration and at $45^{\circ}$ to them in the crossed one.  
For the $B_{g}$ modes, in contrast, the highest intensity is along the crystal axes in the crossed geometry and at the angle of $45^{\circ}$ in the parallel one.  
Thus, this approach allowed us to reliably determine the symmetry of the Raman-active phonons with weak corresponding spectral lines. 
Figure~S4 
shows angular dependences of phonon intensity derived from DFT calculations and a fair agreement with the experiment can be observed (see Fig.~S3 in SM~\cite{supp_mat}).

\begin{figure*}
\centering
\includegraphics[width=2\columnwidth]{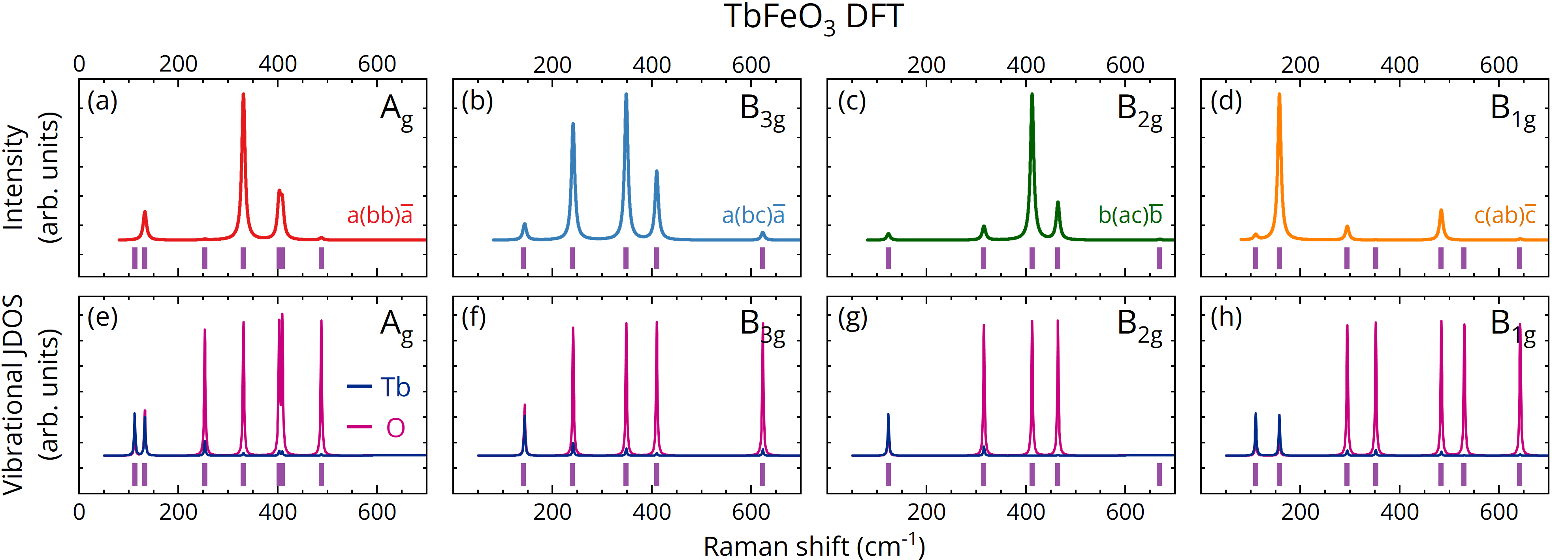}
\caption{\label{fig:raman_dft_pdos}
Calculated polarized Raman spectra and joint density of vibrational states (JDOS) projected onto ionic contributions of (a), (e)~$A_{g}$; (b), (f)~$B_{3g}$; (c), (g)~$B_{2g}$; and (d), (h)~$B_{1g}$ phonons for \TFO, respectively.
The polarization configurations are given in Porto's notation as described in the text.
Purple tick marks at the bottom of each plot present the calculated phonon frequencies.
}
\end{figure*}

\begin{figure}
\centering
\includegraphics[width=1\columnwidth]{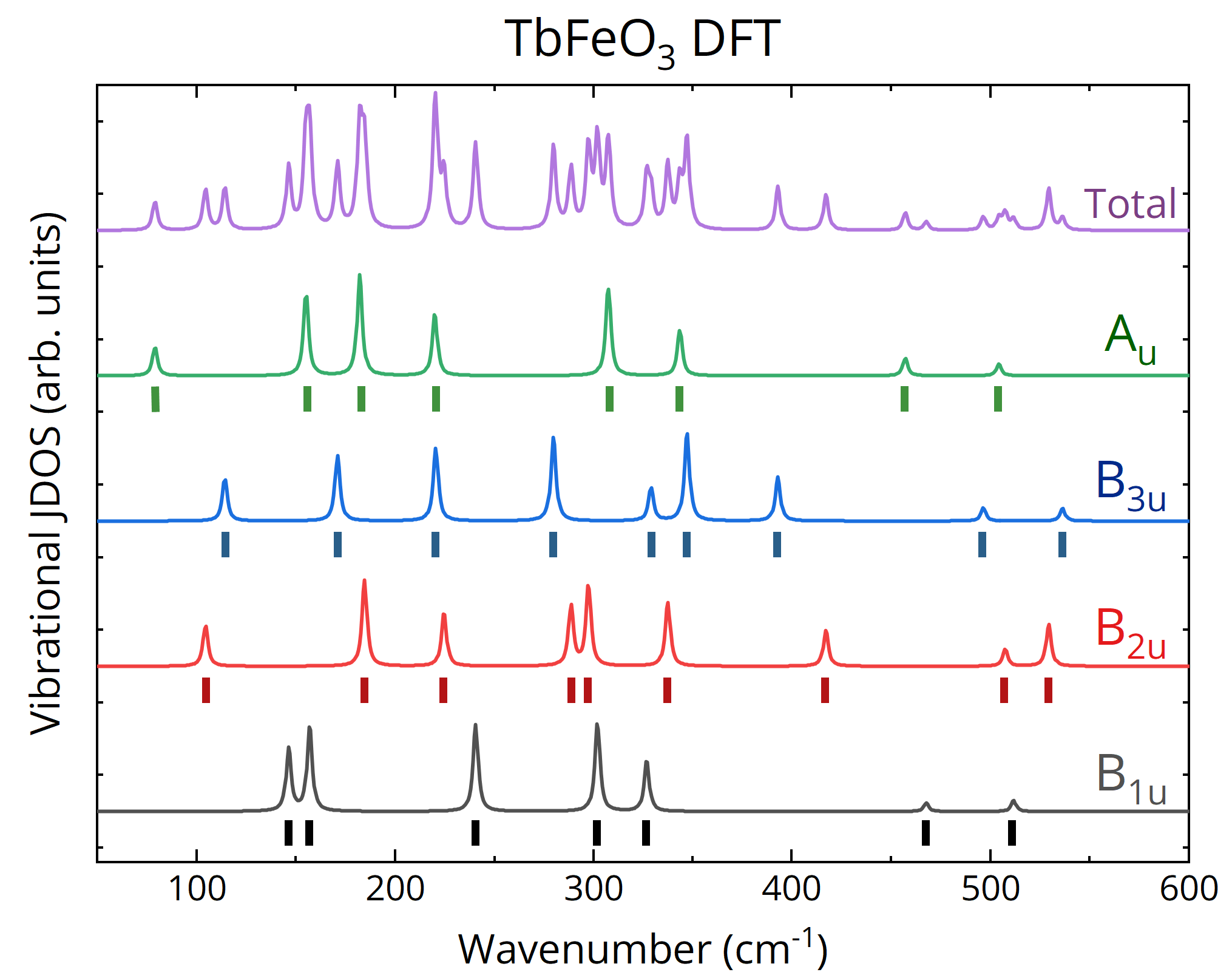}
\caption{\label{fig:Fe_pdos}
Calculated joint density of vibrational states (JDOS) of the Fe ions projected onto the polar $A_{u}$, $B_{1u}$, $B_{2u}$, $B_{3u}$ modes and the total phonon spectrum for \TFO.
Colored tick marks at the bottom of each plot present the calculated phonon frequencies.
}
\end{figure}

Along with polarized Raman spectra [see Figs.~\ref{fig:raman_dft_pdos}(a)-\ref{fig:raman_dft_pdos}(c)], we also have calculated the joint density of vibrational states (JDOS) projected onto ionic contributions for the Raman-active phonons.   
Figures~\ref{fig:raman_dft_pdos}(e)-\ref{fig:raman_dft_pdos}(h) show that Tb and O ions are equally involved in the vibrations corresponding to the low-frequency modes. 
With increasing frequency, due to the large mass of the rare-earth ions, the role of Tb ions in the phonon displacements uniformly decreases and for the vibrations above 480\,cm$^{-1}$ take almost no part.
It should be noted that according to the calculations, the Fe ions remain fixed for all Raman-active phonons.
This is due to the fact that, based on symmetry considerations, the Tb ($4c$) and O ($4c$ and $8d$) ions are active for the both \textit{gerade} $A_{g} \oplus B_{1g} \oplus B_{2g} \oplus B_{3g}$ and \textit{ungerade} $A_{u} \oplus B_{1u} \oplus B_{2u} \oplus B_{3u}$ modes while Fe ($4b$) ions are active only for \textit{ungerade} modes~\cite{kroumova2003bilbao}. 
It follows that for Raman-active modes the involvement of Fe ions is forbidden by symmetry and they can only contribute to polar phonons as shown in Fig.~\ref{fig:Fe_pdos}. 
Apparently, the absence of spin-phonon coupling for Raman-active modes at the magnetic ordering of Fe ions observed in rare-earth orthoferrites is related to this fact~\cite{weber2022emerging}.

\subsection{Lattice dynamics calculations}
\label{subsec:dft}


The common approach to lattice dynamics study within DFT is based on calculations of the dynamical matrix $D$ followed by solving the general eigenvalue problem~\cite{gonze1997dynamical}. 
Taking into account only short-range interaction [analytical (A) contribution] to the dynamical matrix $D^{\alpha\beta}_{ij} = D^{\textrm{A},\alpha\beta}_{ij}$  at the $\Gamma$ point of the Brillouin zone, where $\alpha$ and $\beta$ are the direction indices, and $i$ and $j$ are atomic indices, the solution of the eigenstate equation $D | \xi^{\textrm{TO}}_{m} \rangle = \omega_{\textrm{TO},m}^{2} | \xi^{\textrm{TO}}_{m} \rangle$
gives eigenvector $\xi^{\textrm{TO}}_{m}$ and frequency $\omega_{\textrm{TO},m}$ of the $m$th TO phonon.
To account for the long-range macroscopic electric field which is induced by collective atomic displacements, the nonanalytical (NA) contribution to the dynamical matrix $D^{\alpha\beta}_{ij} = D^{\textrm{A},\alpha\beta}_{ij} + D^{\textrm{NA},\alpha\beta}_{ij}$ is necessary, which in the vicinity of the $\Gamma$ point takes the form~\cite{zhong1994giant,gonze1997dynamical} 
\begin{equation}
\label{eq:dynamical_matrix_NAC}
D^{\textrm{NA},\alpha\beta}_{ij} = \frac{1}{\sqrt{M_{i}M_{j}}}\frac{4 \pi e^2}{\Omega} \frac{[\mathbf{q}\cdot\mathbf{Z}_{i}]_{\alpha}[\mathbf{q}\cdot\mathbf{Z}_{j}]_{\beta}}{\mathbf{q}\cdot\boldsymbol{\varepsilon}_{\infty}\cdot\mathbf{q}}\bigg|_{\mathbf{q}\rightarrow0}, 
\end{equation}
where $M_{i}$ is the mass of the $i$th ion, $e$ is the elementary charge, $\Omega$ is the volume of the unit cell, $\mathbf{Z}_{i}$ is the Born effective charge tensor, $\mathbf{q}$ is the wave vector, and $\boldsymbol{\varepsilon}_{\infty}$ is the high-frequency dielectric permittivity tensor. 
Then the eigenvector $\xi^{\textrm{LO}}_{m}$ and frequency $\omega_{\textrm{LO},m}$ of the $m^{\textrm{th}}$ LO phonon can be obtained by solving the equation $D | \xi^{\textrm{LO}}_{m} \rangle = \omega_{\textrm{LO},m}^{2} | \xi^{\textrm{LO}}_{m} \rangle$.
It is worth noting that the calculations of the TO and LO modes involve the diagonalization of different dynamical matrices $D = D^{\textrm{A}}$ and $D = D^{\textrm{A}} + D^{\textrm{NA}}$, respectively.
Thus, in general the eigenvectors $\xi^{\textrm{LO}}$ and $\xi^{\textrm{TO}}$ for polar phonons are not necessarily equal.
Moreover, the NA term $D^{\textrm{NA},\alpha\beta}_{ij}$ is nondiagonal and often causes a strong mixing of different modes due to the Coulomb interaction; i.e., several TO modes may contribute to a single LO mode~\cite{raeliarijaona2015mode,fredrickson2016theoretical}. 
Note that the NA contribution affects only polar phonons, whereas the frequencies and eigenvectors of nonpolar phonons remain unchanged in the vicinity of the $\Gamma$ point.


It is well known that each normal (TO) mode corresponds to a single irreducible representation of the point group of the crystal, whereas all other possible modes can be represented as linear combinations of these irreducible representations~\cite{dove1993introduction}.
That is, the eigenvectors of normal modes satisfy the orthonormal conditions $\langle \xi^{\textrm{TO}}_{m} | \xi^{\textrm{TO}}_{n} \rangle = \delta_{mn}$, where $\delta_{mn}$ is the Kronecker delta, and form a basis.
Thus, it is reasonable to expand the eigenvector of the $m$th LO mode to a linear combination of the normal modes,
\begin{equation}
\label{eq:decomposition_LOTO}
| \xi^{\textrm{LO}}_{m} \rangle = \sum_{n} C_{mn} | \xi^{\textrm{TO}}_{n} \rangle.    
\end{equation}
Note that non-zero expansion coefficients $C_{mn} \neq 0$ can give only polar modes with the same polarization (basis function).
The contribution from the acoustic mode is usually negligible, so only the polar TO modes are taken into account in the decomposition from Eq.~\eqref{eq:decomposition_LOTO}.
Thereby, it is useful to analyze the overlap matrix which represents the degree of correlation between the $m^{\textrm{th}}$ LO and $n^{\textrm{th}}$ TO eigenvectors of polar phonons with a specific symmetry according to the expression~\cite{zhong1994giant,ratnaparkhe2020calculated}
\begin{equation}
\label{eq:correlation_LOTO}
C_{mn} = \langle \xi^{\textrm{LO}}_{m} | \xi^{\textrm{TO}}_{n} \rangle,
\end{equation}
where $\langle\ldots\rangle$ denotes scalar product.
When the TO-LO rule is strictly satisfied for all polar phonons, the $C$ matrix takes a form in which the elements on the main diagonal are manyfold larger than the others.
For ideal crystals where eigenvectors for TO and LO modes are equal $| \xi^{\textrm{TO}} \rangle = | \xi^{\textrm{LO}} \rangle$ the overlap matrix $C$ is the identity matrix with ones on the main diagonal and zeros elsewhere. 
In real crystals, mode mixing caused by the Coulomb interaction is expressed in the form when for some LO modes the relevant elements of the overlap matrix $C$ are essentially non-zero for several TO modes~\cite{zhong1994giant,lee1994lattice,khedidji2021microscopic}.

\begin{table*}
\caption{\label{tab:Born_charges} Calculated Born effective charge tensors $\mathbf{Z}$ of the Tb, Fe and O ions in \TFO.}
\begin{ruledtabular}
\begin{tabular}{cccc}
Tb ($4c$) & Fe ($4b$) & O ($4c$) & O ($8d$) \\ \hline 
$\begin{pmatrix}
             4.02 & 0.29 & 0 \\ 
             0.2  & 3.98 & 0 \\
             0    & 0    & 3.6
\end{pmatrix}$ &
$\begin{pmatrix}
              4.02 & 0.39 & 0.48 \\ 
             -0.19 & 4.25 & -0.12 \\ 
             -0.32 & -0.2 & 4.03
\end{pmatrix}$ &
$\begin{pmatrix}
              -2.55 & -0.32 & 0 \\ 
              -0.46 & -2.12 & 0 \\
                  0 &  0    & -3.32
\end{pmatrix}$ &
$\begin{pmatrix}
             -2.75 & -0.53 & -0.08 \\ 
             -0.57 & -3.05 & -0.17 \\
             -0.04 & -0.2  & -2.16
\end{pmatrix}$
\end{tabular}
\end{ruledtabular}
\end{table*}

\begin{table}
\caption{\label{tab:DTF_IR_phonons} Calculated frequencies $\omega_{\textrm{TO}}$, $\omega_{\textrm{LO}}$, $\tilde \omega_{\textrm{LO}}$ (cm$^{-1}$) and oscillator strengths $S$ (cm$^{-2}$) of the polar $B_{3u}$, $B_{2u}$, and $B_{1u}$ phonons for \TFO.}
\begin{ruledtabular}
\begin{tabular}{ccccc}
Sym. & $\omega_{\textrm{TO}}$ & $\omega_{\textrm{LO}}$ & $\tilde \omega_{\textrm{LO}}$ & $S$ \\ \hline
\multirow{9}{*}{$B_{3u}$} & 112.4 & 112.8 & 115.3 & 4.3~10$^{3}$ \\
                          & 166.7 & 175.6 & 225.3 & 1.4~10$^{5}$ \\
                          & 230.3 & 266.9 & 412.1 & 6.8~10$^{5}$ \\
                          & 274.7 & 314.1 & 307.6 & 1.1~10$^{5}$\\
                          & 322.5 & 323   & 323.1 & 2.2~10$^{3}$ \\ 
                          & 341.4 & 361.1 & 376.5 & 1.5~10$^{5}$\\
                          & 390.7 & 490.9 & 464.6 & 3.6~10$^{5}$\\ 
                          & 495   & 507.3 & 496.8 & 1.1~10$^{4}$ \\
                          & 521.5 & 610.8 & 531.3 & 6.1~10$^{5}$ \\ \hline
\multirow{9}{*}{$B_{2u}$} & 101.3 & 101.8 & 103.7 & 3~10$^{3}$ \\
                          & 182.3 & 182.4 & 182.4 & 2~10$^{2}$ \\
                          & 226.2 & 249.8 & 368.6 & 5.1~10$^{5}$\\
                          & 286.5 & 287.5 & 315.5 & 1.1~10$^{5}$\\
                          & 294.6 & 396.3 & 465.6 & 7.7~10$^{5}$\\
                          & 329.6 & 329.6 & 329.6 & 25 \\
                          & 416.8 & 480.4 & 450.4 & 1.7~10$^{5}$\\
                          & 491.8 & 513.3 & 502.6 & 6.3~10$^{4}$  \\
                          & 514.4 & 614.6 & 515.6 & 7.7~10$^{3}$  \\ \hline
\multirow{7}{*}{$B_{1u}$} & 149.4 & 153.1 & 217.1 & 1.5~10$^{5}$\\ 
                          & 155.7 & 174.3 & 192.9 & 6.4~10$^{4}$ \\
                          & 243.7 & 284.6 & 377.0 & 4.7~10$^{5}$\\
                          & 297.2 & 297.3 & 297.3 & 3.5~10$^{2}$   \\
                          & 332.2 & 442.4 & 455.0 & 5.6~10$^{5}$\\
                          & 459.3 & 472.8 & 467.6 & 4.3~10$^{3}$ \\
                          & 491.4 & 600.3 & 512.1 & 1.2~10$^{5}$\\
\end{tabular}
\end{ruledtabular}
\end{table}

To gain insight into the phonon landscape of the orthoferrite \TFO, we performed the first principles calculations of the lattice dynamics in the vicinity of the $\Gamma$ point of the Brillouin zone.
The calculated lattice parameters $a$, $b$, and $c$, static $\varepsilon_{0}$ and high frequency $\varepsilon_{\infty}$ anisotropic dielectric permittivities in comparison to experimental values are listed in Table~\ref{tab:epsilon}.
The Born effective charge tensors $\mathbf{Z}$ of Tb, Fe and O ions in \TFO{} are listed in Table~\ref{tab:Born_charges}. 
The LO modes were obtained with the NA term [Eq.~\eqref{eq:dynamical_matrix_NAC}] taken into account in the calculations.  
The computed frequencies $\omega_{\textrm{TO}}$ and $\omega_{\textrm{LO}}$ and dielectric strengths $\Delta\varepsilon$ of the polar phonons are listed in parentheses in Table~\ref{tab:IR_phonons}.
Note that there is a fair agreement between the calculation and experimental results. 
Moreover, the obtained frequencies of the Raman-active phonons are in good agreement with experimental data presented in Refs.~\cite{venugopalan1985magnetic,weber2016raman} as can be seen in Table~\ref{tab:Raman_phonons}.


\begin{figure*}
\centering
\includegraphics[width=2\columnwidth]{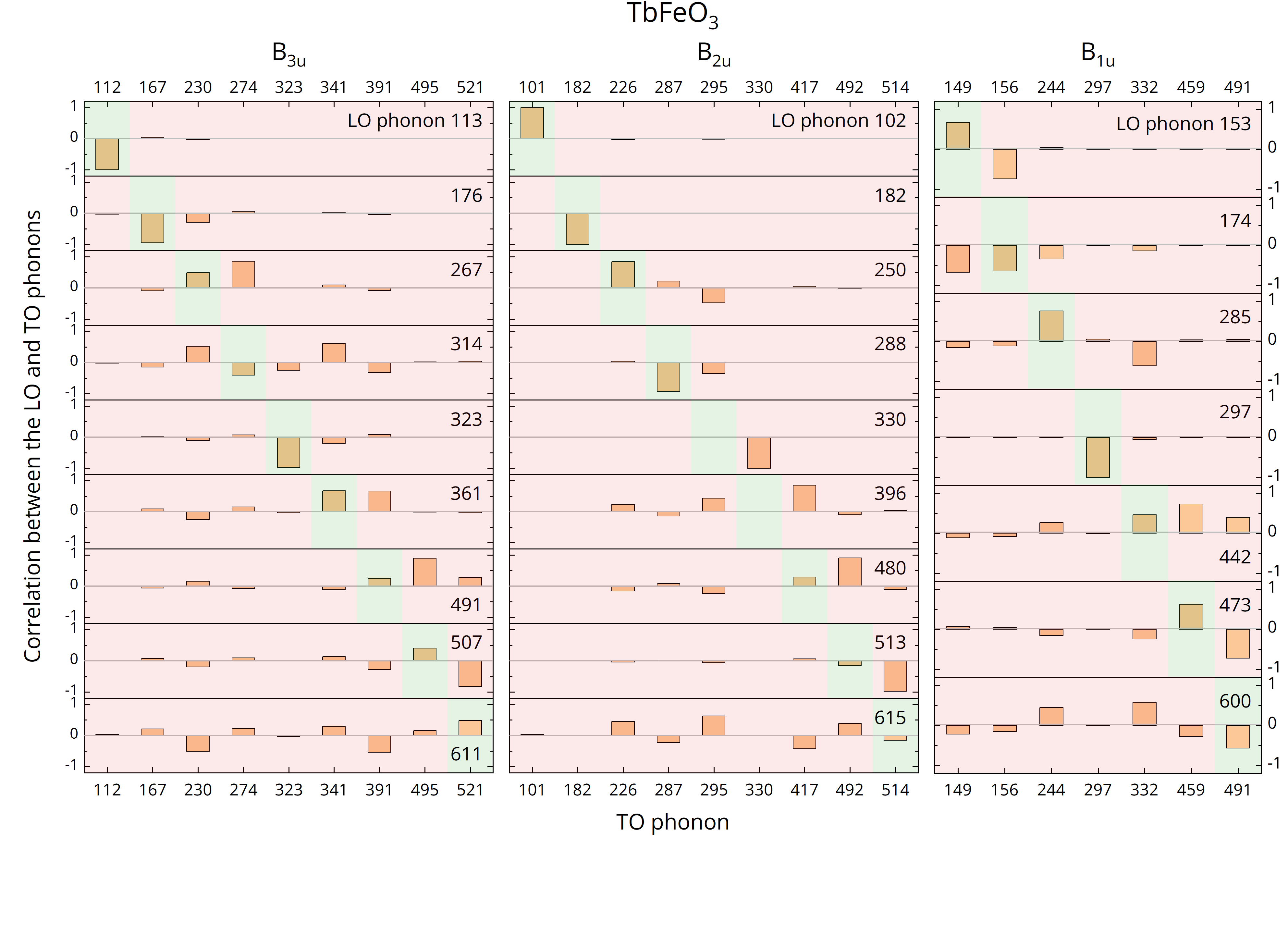}
\caption{\label{fig:LOTO_correlation}
The overlap matrices $C$ which represent correlations between eigenvectors of LO and TO polar phonons with $B_{3u}$ (left panel), $B_{2u}$ (center panel), and $B_{1u}$ (right panel) symmetry according to the DFT calculations at the $\Gamma$ point of the Brillouin zone in \TFO.  
The values of the LO phonon frequencies are given.
The green background corresponds to the area where the ``TO-LO rule'' should be satisfied.
}
\end{figure*}

To establish the relationship between the TO and LO modes, we calculated the overlap matrices $C$ which represent the correlations between their eigenvectors using Eq.~\eqref{eq:correlation_LOTO}. 
The resulting overlap matrices $C$ for $B_{3u}$, $B_{2u}$, and $B_{1u}$ polar phonons are presented as a bar chart in Fig.~\ref{fig:LOTO_correlation}.
Here, the green background highlights the main diagonal of the overlap matrix $C$ where the ``TO-LO rule'' should be satisfied.
In other words, if this rule is met, the bar in the green background must be many times greater then those in the red background.  
Fig.~\ref{fig:LOTO_correlation} clearly shows that the ``TO-LO rule'' is strictly fulfilled only for a few LO modes with frequencies 113\,cm$^{-1}$, 176\,cm$^{-1}$, 323\,cm$^{-1}$ for $B_{3u}$, 102\,cm$^{-1}$, 182\,cm$^{-1}$, 288\,cm$^{-1}$ for $B_{2u}$, and 297\,cm$^{-1}$ for $B_{1u}$.
For other LO modes (e.g. 491\,cm$^{-1}$, 507\,cm$^{-1}$ for $B_{3u}$, and 480\,cm$^{-1}$ for $B_{2u}$) the relevant TO modes identified from the correlation analysis have a higher frequency so that $\omega_{\textrm{TO}} > \omega_{\textrm{LO}}$, thus breaking the ``TO-LO rule''.

The highest frequency LO modes (611\,cm$^{-1}$ for $B_{3u}$, 615\,cm$^{-1}$ for $B_{2u}$, and 600\,cm$^{-1}$ for $B_{1u}$,) largely correspond to the several lower-frequency TO modes due to the mixing described above as can be seen in Fig.~\ref{fig:LOTO_correlation}.
Perhaps this pronounced mixing manifests itself in the nonlinear magneto-phononic effects observed for the highest frequency LO modes in orthoferrites.
Thus, the resonance mid-infrared pumping of these LO modes causes coherent spin and lattice dynamics at the frequencies of the quasi-antiferromagnetic resonance (25\,cm$^{-1}$) and $A_{g}$ modes (112\,cm$^{-1}$ and 162\,cm$^{-1}$) in the rare-earth orthoferrite $\mathrm{ErFeO}_{3}$~\cite{nova2017effective,juraschek2017ultrafast,venugopalan1985magnetic}.     
The counter intuitive result here is that the direct excitation of LO modes by a transverse electromagnetic wave in bulk material should be forbidden because $\varepsilon_{2}(\omega_{\textrm{LO}}) = 0$ assuming that $\gamma_{\textrm{LO}} = 0$~\cite{schrader2008infrared}. 
Furthermore, the mechanism of nonlinear coupling between the high-frequency polar LO modes and Raman-active $A_{g}$ modes was not disclosed in Ref.~\cite{nova2017effective}.
It is worth noting that the Raman-active $A_{g}$ modes and polar phonons have the same symmetry away from the $\Gamma$ point and thereby can directly interact with each other in the Brillouin zone, as discussed below.
\textcolor{newtext}{Moreover, LO modes in crystals have attracted special attention due to recently observed strongly enhanced light-matter interaction in the phononic epsilon-near-zero regime $\varepsilon_{2}(\omega_{\textrm{LO}}) = 0$ which allows to switch the spin and polarization order parameters~\cite{stupakiewicz2021ultrafast,kwaaitaal2024epsilon,davies2024epsilon}.}



\subsection{LO-TO mixing}
\label{subsec:LOTO}

\begin{figure*}
\centering
\includegraphics[width=2\columnwidth]{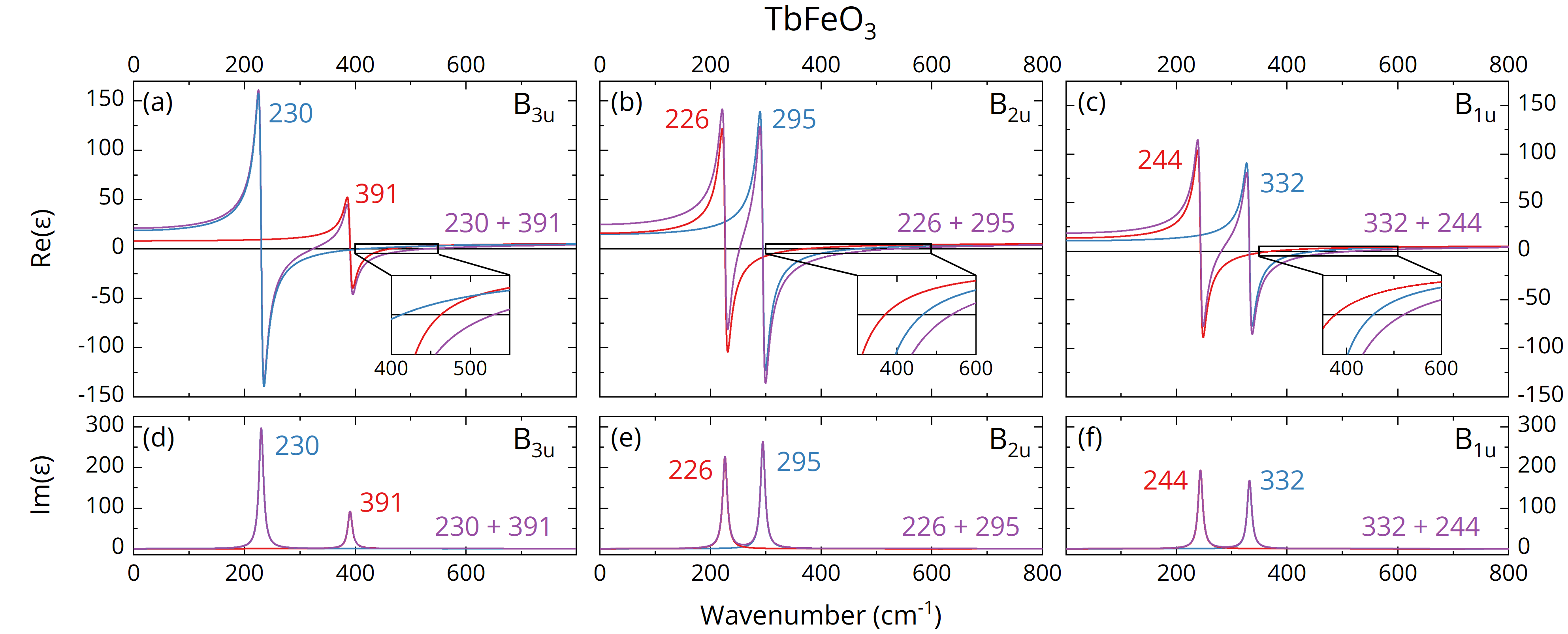}
\caption{\label{fig:eps12}
Calculated spectra of the real and imaginary parts of the dielectric permittivity $\varepsilon(\omega)$ 
of the (a), (d) $B_{3u}$, (b), (e) $B_{2u}$ and (c), (f) $B_{1u}$ polar phonons with the highest mode dynamical charges, respectively. 
The frequencies of the considered polar TO phonons are given.
}
\end{figure*}

To disclose how mixing of the polar TO phonons affects the LO modes and enables excitation of the latter by electromagnetic waves, we consider the effect of mode dynamical charges on the spectra of the complex dielectric permittivity $\varepsilon(\omega)$ in \TFO.    
Note that the highest frequency LO modes of different symmetries correspond to the major peaks in the spectra of the imaginary part of the inverse dielectric permittivity $-\Im[\varepsilon^{-1}(\omega)]$ as shown in Figs.~\ref{fig:reflectivity}(g)--\ref{fig:reflectivity}(i).
Furthermore, these LO modes have the \textcolor{newtext}{evident} correlation with most TO phonons as can be seen in Fig.~\ref{fig:LOTO_correlation}. 
Thus, it is convenient to analyze the relationship of the highest frequency LO modes of different symmetries with the TO modes \textcolor{newtext}{of the same symmetry}.
For this, we consider the complex dielectric functions induced by some TO modes according to the classical Lorentz oscillator model~\cite{gonze1997dynamical} 
\begin{equation}
\label{eq:epsilon_harmonic}
\varepsilon(\omega) = \varepsilon_{\infty} + \frac{4\pi}{\Omega} \sum_{j}\frac{S_{j}}{{\omega^{2}_{j\textrm{TO}}} - {\omega}^2 + i\gamma_{j\textrm{TO}}\omega},
\end{equation}
where $S_{j}$ is the oscillator strength (or mode dynamical charge) of the $j$th polar TO phonon determined by equation
\begin{equation}
\label{eq:mode_charge}
S_{j,\alpha} = \left( \sum_{i,\beta} \frac{1}{\sqrt{M_i}}\xi_{i,j\beta}Z^{*}_{i,\alpha\beta} \right) ^2,
\end{equation}
where $Z^{*}_{i}$ is the Born effective charge of the $i^{\textrm{th}}$ atom, respectively and $\xi_{i,j}$ is the $i^{\textrm{th}}$ component of the dynamical matrix $j^{\textrm{th}}$ eigenvector, and the other parameters have the same meaning as in Eq.~(S1) in SM~\cite{supp_mat}. 
It is worth noting that this classical model [Eq.~\eqref{eq:epsilon_harmonic}] can be reduced to the factorized form of the dielectric permittivity [Eq.~(S1) in SM~\cite{supp_mat}]
under the assumption of equality of the dampings $\gamma_{j\textrm{LO}} = \gamma_{j\textrm{TO}}$.
The calculated oscillator strengths $S$, phonon frequencies $\omega_{\textrm{TO}}$, and high-frequency dielectric permittivity $\varepsilon_{\infty}$ are listed in Tables~\ref{tab:DTF_IR_phonons} and~\ref{tab:epsilon}, respectively.  



We now consider the model with a single polar mode, which means that the remaining phonon modes of the same symmetry are excluded from the analysis.
In this model the eigenvectors of the TO and LO modes at the $\Gamma$ point of the Brillouin zone are identical, $| \xi^{\textrm{LO}} \rangle = | \xi^{\textrm{TO}} \rangle$.
Then the following relationship for phonon frequencies holds~\cite{lee1994lattice,gonze1997dynamical}   
\begin{equation}
\label{eq:longitudinal_omega_tilde}
\tilde \omega^{2}_{m\textrm{LO}} = \omega^{2}_{m\textrm{TO}} + \frac{4 \pi}{\Omega} \frac{{\mathbf{q}\cdot\mathbf{S}_{m}\cdot\mathbf{q}}}{\mathbf{q}\cdot\boldsymbol{\varepsilon}_{\infty}\cdot\mathbf{q}}\bigg|_{\mathbf{q}\rightarrow0},
\end{equation}
where $\tilde \omega_{m\textrm{LO}}$ is the frequency of the LO phonon within a single polar mode model.
The evaluated $\tilde \omega_{m\textrm{LO}}$ values compared to results of calculations using a ``real'' dynamical matrix with a nonanalytical term included in Eq.~\eqref{eq:dynamical_matrix_NAC} are listed in Table~\ref{tab:DTF_IR_phonons}.
From the discrepancy between $\omega_{m\textrm{LO}}$ and $\tilde \omega_{m\textrm{LO}}$ one can readily conclude that in real crystals the LO phonon states correspond to a complex set of TO polar vibrational modes.


\begin{figure*}
\centering
\includegraphics[width=2\columnwidth]{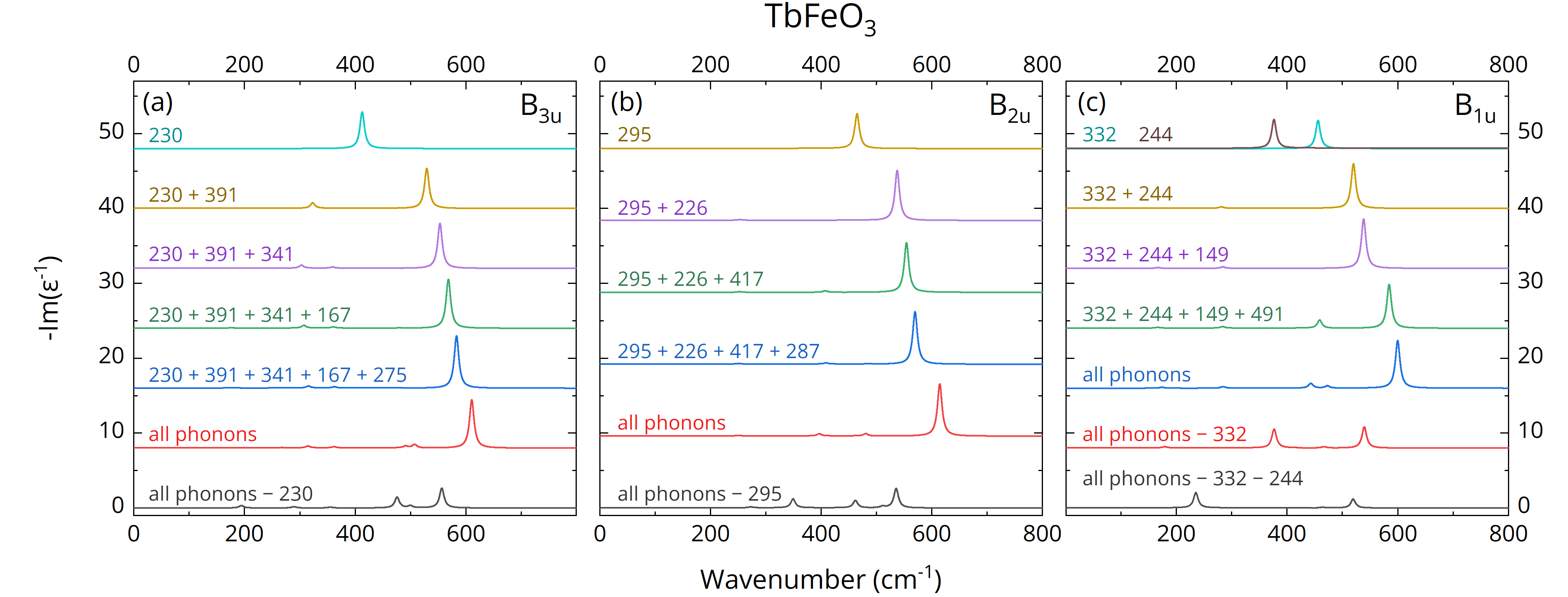}
\caption{\label{fig:inveps}
Spectra of the imaginary part of the inverse dielectric permittivity $-\Im[\varepsilon^{-1}(\omega)]$ with contributions from the (a)~$B_{3u}$, (b)~$B_{2u}$, and (c)~$B_{1u}$ polar LO phonons obtained from results of the DFT calculation for \TFO{} using Eq.~\eqref{eq:epsilon_harmonic}.
The effect of strong mixing of polar TO phonons with high dynamical charges on the highest frequency LO mode which corresponds to the major peak is shown.  
The frequencies of the considered polar TO phonon are given.
}
\end{figure*}

In order to establish the influence of polar TO phonons on LO vibrational states one can simulate the spectrum of $-\Im[\varepsilon^{-1}(\omega)]$ which reveals, as mentioned above, the peculiarities at frequencies of LO phonons.
The spectrum is simulated using partial summation in Eq.~\eqref{eq:epsilon_harmonic} and plotted in Fig.~\ref{fig:inveps}.
As an initial approximation, the spectrum was simulated using a single polar model (top curves in Fig.~\ref{fig:inveps}) by taking into account only TO modes with the highest values of oscillator strength $S$, namely 230\,cm$^{-1}$ for $B_{3u}$, 295\,cm$^{-1}$ for $B_{2u}$, and 332 and 244\,cm$^{-1}$ for $B_{1u}$ (Table~\ref{tab:DTF_IR_phonons}).
These modes give the most significant impact in comparison to all other modes in the spectra $\Im[\varepsilon(\omega)]$ and $-\Im[\varepsilon^{-1}(\omega)]$ (see Fig.~\ref{fig:eps12}).
Then, by including additional modes in the sum of Eq.~\eqref{eq:epsilon_harmonic} the spectrum evolution can be clearly seen as the gradual increase in intensity and frequency of the highest frequency band in the spectra as seen in Fig.\ref{fig:inveps}.
This effect stems from the fact that individual oscillators modify the complex dielectric permittivity significantly beyond their resonance TO frequency.
Moreover, upon exclusion of TO phonons with the highest oscillator strength, the major high frequency band vanishes as can be seen in Fig.~\ref{fig:inveps} (bottom row).
Thus these TO phonons also have the strongest correlation with the highest frequency LO modes (see Fig.~\ref{fig:LOTO_correlation}).
Therefore, using the highest-frequency LO modes as an example, we have demonstrated that the correlation between the LO and TO modes is due to the mixing of harmonic (uncoupled) TO phonons, which reproduces the results obtained using the overlap matrix technique shown in Fig.~\ref{fig:LOTO_correlation}.

The obtained results can be summarized as a multimode model with polar modes taking the role of dynamical charges with an exciting long-range electric field, the strength of which is proportional to the dynamical charges of polar phonons.
Then the frequency, modulation, and strength of the field determine the LO states.
This explains the complex nature of LO modes in the real crystals with several polar mode vibrational states.
It is interesting to note that, in a similar way, on the example of the two harmonic oscillators, it is possible to show a breaking of the TO-LO rule. 
The analysis of the spectrum of the imaginary part of the inverse dielectric permittivity shows that the mode mixing between one phonon with a high mode dynamical charge characterized by a strong LO-TO splitting and another phonon with a small mode dynamical charge and a weak LO-TO splitting inside the first one leads to the frequency inversion $\omega_{\textrm{LO}} < \omega_{\textrm{TO}}$ of the later phonon~\cite{fredrickson2016theoretical}.
Furthermore, this effect is only caused by the mode mixing because both phonons have dynamical charges of the same sign. 

\begin{figure*}
\centering
\includegraphics[width=2\columnwidth]{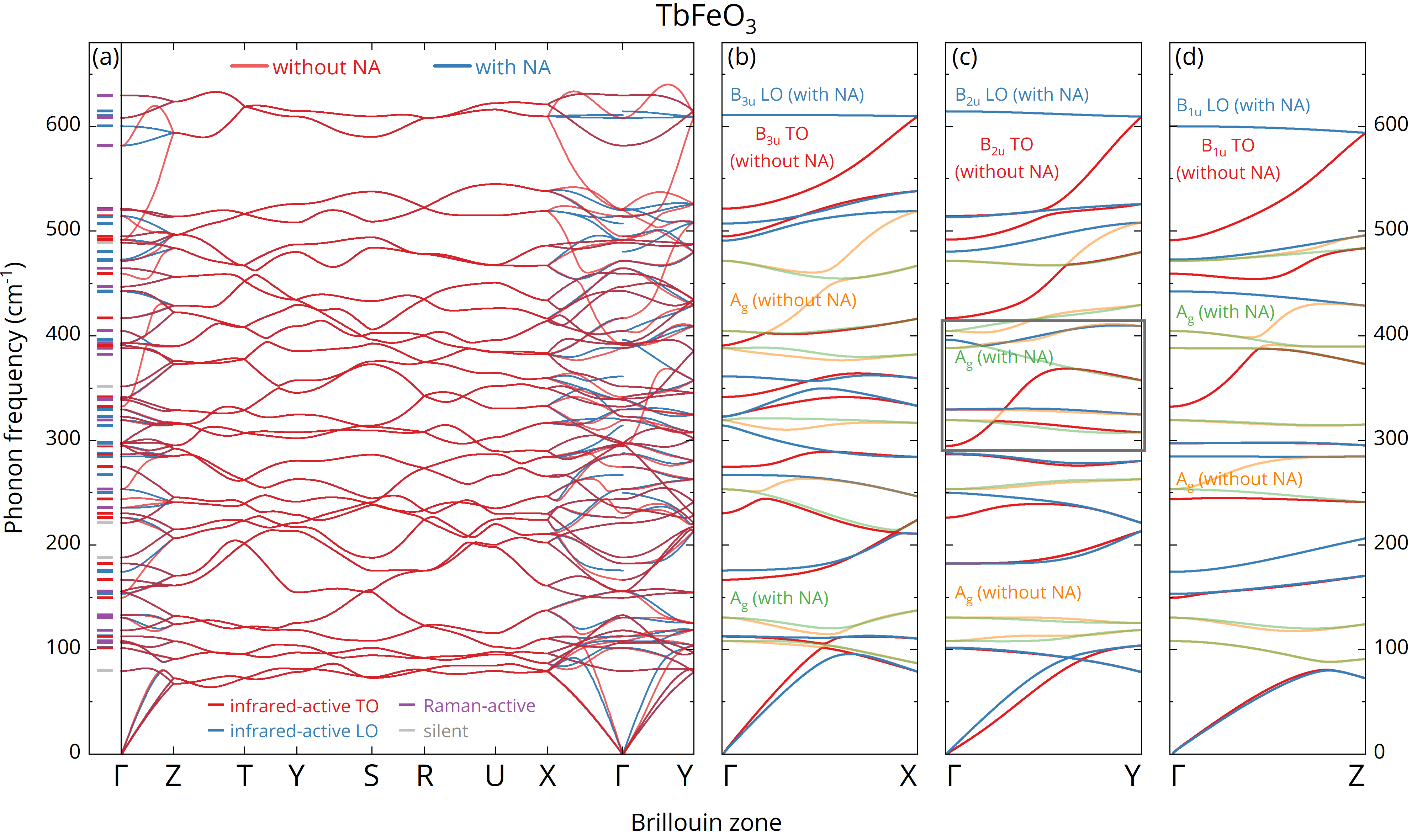}
\caption{\label{fig:phonon_dispersion}
(a)~Full phonon dispersion curves along the $\Gamma$--$\mathrm{Z}$--$\mathrm{T}$--$\mathrm{Y}$--$\mathrm{S}$--$\mathrm{R}$--$\mathrm{U}$--$\mathrm{X}$--$\Gamma$--$\mathrm{Y}$ high-symmetry path of the Brillouin zone for \TFO{} according to the lattice dynamics calculations with and without the NA term.
Dispersion curves for phonons with the same symmetry along the (a)~$\Gamma$--$\mathrm{X}$, (b)~$\Gamma$--$\mathrm{Y}$, and (c)~$\Gamma$--$\mathrm{Z}$ paths calculated with and without the NA term.
The gray frame marks the conditions for the negative LO-TO splitting.
\textcolor{newtext}{Colored tick marks on the left side of the panel (a) present the calculated phonon frequencies at the $\Gamma$ point.}
}
\end{figure*}

\begin{figure*}
\centering
\includegraphics[width=2\columnwidth]{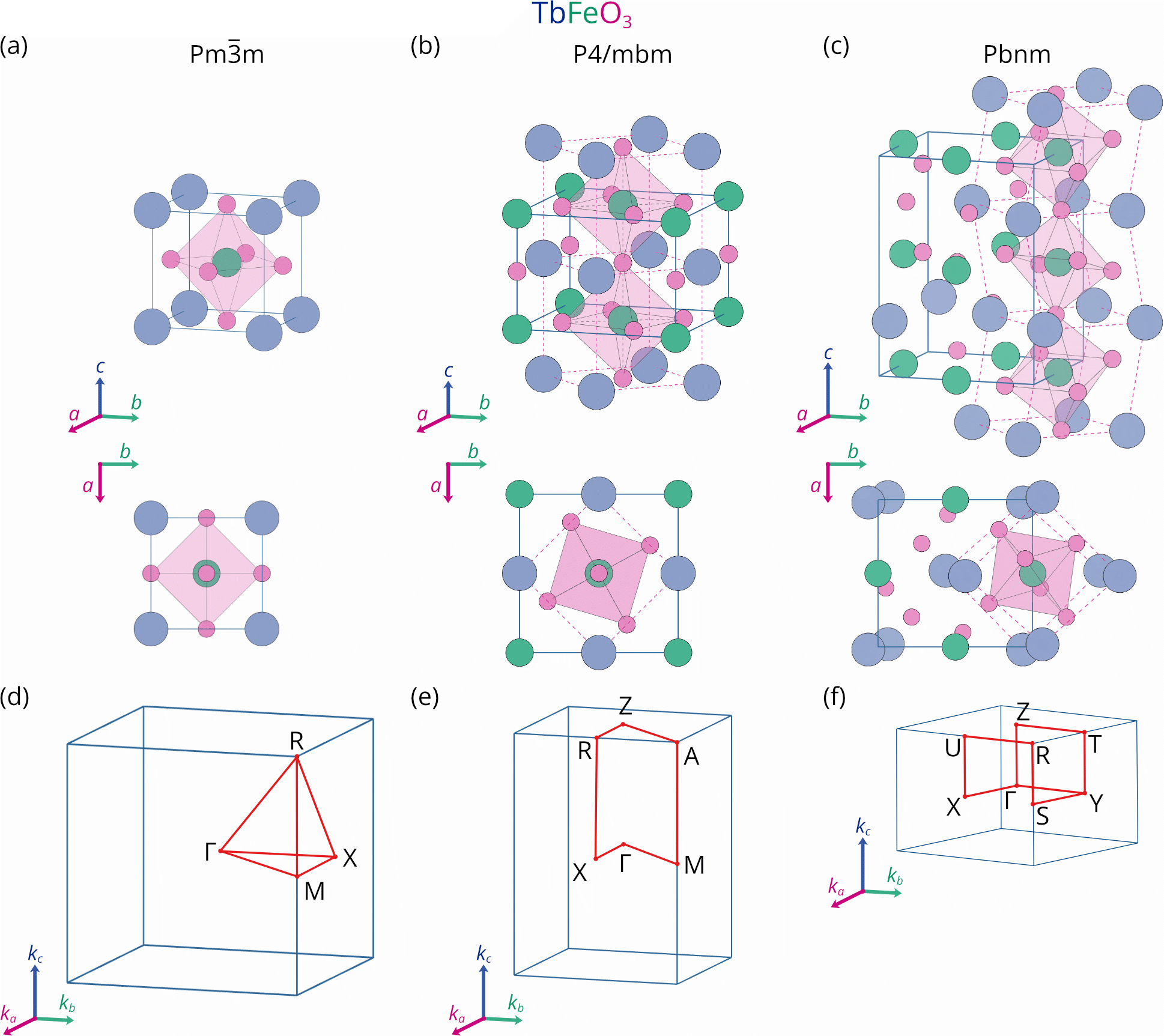}
\caption{\label{fig:structure}
Crystal structure of the rare-earth orthoferrite \TFO{} in the (a)~cubic $Pm\overline{3}m$, (b)~tetragonal $P4/mbm$, and (c)~orthorhombic $Pbnm$ phases. 
Distortions of the ideal cubic perovskite structure are shown by dashed lines.
First Brillouin zone of (d)~cubic, (e)~tetragonal, and (f)~orthorhombic lattice indicating high-symmetry points and paths used in the lattice dynamics simulations. 
The $k_{a}$, $k_{b}$, and $k_{c}$ are the primitive reciprocal lattice vectors.
Visualizations of the crystal structures were prepared using the \textsc{VESTA} software~\cite{momma2011vesta}.
}
\end{figure*}

To reveal an unambiguous way in association of the LO modes with the TO ones we performed the lattice dynamical calculations of \TFO{} with and without the NA term along the high-symmetry paths of the Brillouin zone represented in Fig.~\ref{fig:structure}(b).
The obtained dispersion curves of phonons are shown in Fig.~\ref{fig:phonon_dispersion}(a).
It is clearly seen that taking the NA term into account alters the dispersion curves of only some phonons originating from the $\Gamma$ point of the Brillouin zone.
These dispersion curves at the $\Gamma$ point correspond to phonons with the symmetry $B_{3u}$, $A_{g}$ for $\Gamma$--$\mathrm{X}$, $B_{2u}$, $A_{g}$ for $\Gamma$--$\mathrm{Y}$, and $B_{1u}$, $A_{g}$ for $\Gamma$--$\mathrm{Z}$ paths, as shown in Figs.~\ref{fig:phonon_dispersion}(b)--\ref{fig:phonon_dispersion}(d).
Note that there are an anticrossings between the modes of the same symmetry, present in both cases, with and without the NA term.
Furthermore, for most dispersion curves of phonons with the same symmetry the inclusion of the NA term changes the dispersion close to the $\Gamma$ point, while this effect vanishes at the edge of the Brillouin zone 
because of the long range character of the Coulomb interaction. 
Thus, analyzing these dispersion curves merged at the boundary of the Brillouin zone, one can reliably associate most of the LO and TO modes to each other also at the $\Gamma$ point.
The challenging case arises when dispersion curves calculated with and without the NA term have different symmetries at the $\Gamma$ point, e.g. $B_{3u}$ and $A_{g}$, and merge at the boundary of the Brillouin zone.


Through this elaborate analysis of the phonon dispersion curves, it was revealed that, despite the complex form of the correlation matrix $C$ (see Fig.~\ref{fig:LOTO_correlation}), the ``LO-TO'' rule is not fulfilled for only two polar phonons in \TFO.
Specifically, the $B_{2u}$ mode with a calculated TO frequency $\omega^{S}_{\textrm{TO}} = 294.6$\,cm$^{-1}$ has the corresponding LO frequency $\omega^{S}_{\textrm{LO}} = 396.3$\,cm$^{-1}$.
This LO mode with strong (S) LO-TO splitting leapfrogs the $B_{2u}$ mode with a frequency 329.6\,cm$^{-1}$ and a very weak (W) LO-TO splitting as shown in Fig.~\ref{fig:phonon_dispersion}(c).
In this case, the mode mixing leads to permutation of the TO and LO frequencies and a negative LO-TO splitting $\omega^{W}_{\textrm{LO}} < \omega^{W}_{\textrm{TO}}$ occurs according to the theory from Ref.~\cite{fredrickson2016theoretical}.
Moreover, since the phonon dispersion curves of the same symmetry do not cross, the existence of a polar phonon with a negative LO-TO splitting in orthoferrites requires the presence of at least one $A_{g}$ mode with frequency between $\omega^{S}_{\textrm{TO}} < \omega_{A_{g}} < \omega^{S}_{\textrm{LO}}$ as shown in the grey frame in Fig.~\ref{fig:phonon_dispersion}(c).  
It is worth noting that the negative LO-TO splitting of this phonon is also observed in our experimental results shown in Fig.~\ref{fig:reflectivity}(b). 
The assignment of the calculated frequencies of the TO and LO modes allowed us to connect the TO and LO modes obtained in the experiment as listed in Table~\ref{tab:IR_phonons}. 
Therefore, the presented analysis of the calculated phonon dispersion curves allowed us to consistently and unambiguously associate the corresponding TO and LO modes with each other in experimental spectra for \TFO.


\subsection{Phonon genesis}
\label{subsec:genesis}

In order to complete the analysis of the phonon states, we establish a genetic relationship between phonons in the orthorhombic and parent cubic phases using group theory.
It is known that several paths from the parent cubic $Pm\overline{3}m$ to orthorhombic $Pbnm$ phase for perovskites are possible~\cite{aleksandrov1976sequences,kroumova2003bilbao}.
Among them a sequence of two transformations $Pm\overline{3}m \xleftrightarrow{\textrm{1st}} I4/mcm \xleftrightarrow{\textrm{2nd}} Pbnm$ with the first- and second-order phase transitions, respectively, was experimentally confirmed in the perovskite~\cite{ali2005space}.
However, the first-order transition considerably obstructs a joint analysis of the lattice dynamics in related phases and revealing the connection between phonons \textcolor{newtext}{because the lattice parameters and therefore phonon frequencies change abruptly breaking the connection between modes in different phases}.
On the other hand, the $Pm\overline{3}m \xleftrightarrow{\textrm{2nd}} P4/mbm \xleftrightarrow{\textrm{2nd}} Pbnm$ path is symmetry-allowed and we use this path in our analysis \textcolor{newtext}{as it allows us to establish a well-defined relation between phonons in different phases}~\cite{kroumova2003bilbao,wang2021finite}.
It is worth noting that the tetragonal $P4/mbm$ structure is realized in perovskite crystals~\cite{martin2005effect,fabini2016reentrant}.

\begin{figure}
\centering
\includegraphics[width=1\columnwidth]{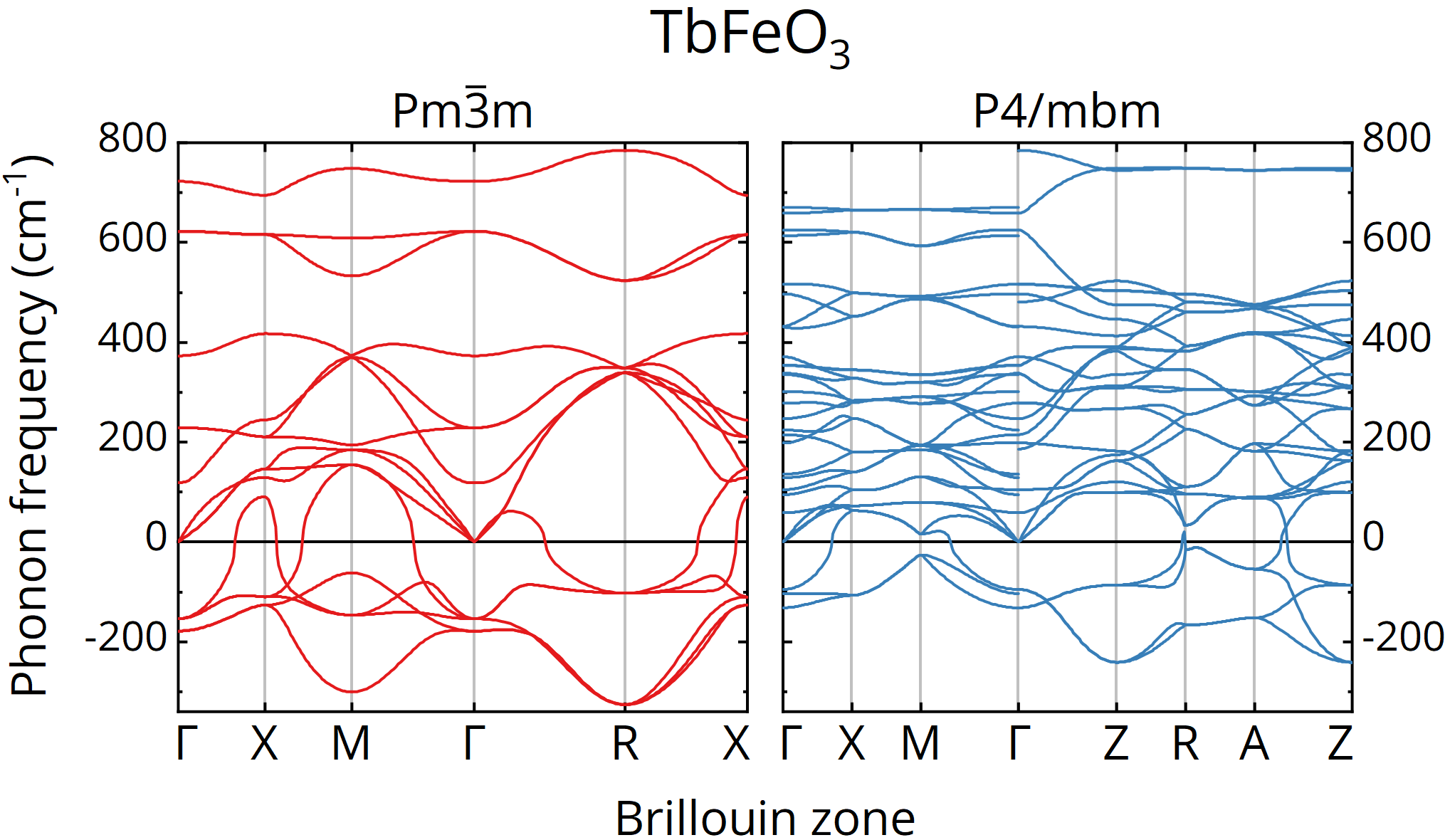}
\caption{\label{fig:DFT_Pm3m_P4mbm}
Full phonon dispersion curves calculated for the (a)~cubic $Pm\overline{3}m$ and (b)~tetragonal $P4/mbm$ phases of the rare-earth orthoferrite \TFO{}.
}
\end{figure}

\begin{figure}
\centering
\includegraphics[width=1\columnwidth]{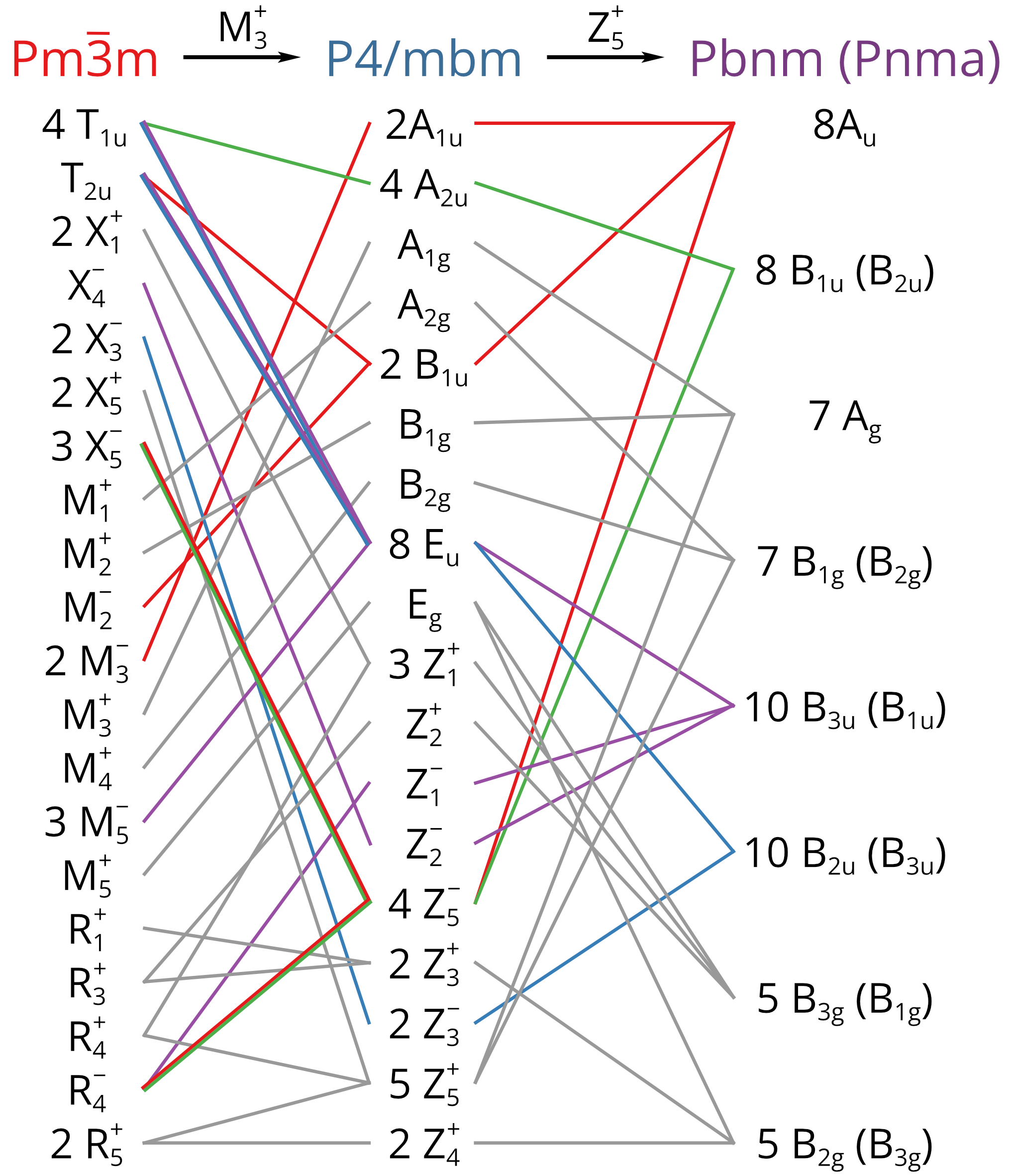}
\caption{\label{fig:phcorrel}
Correlation diagram between the space group representations for parent cubic $Pm\overline{3}m$ and orthorhombic $Pbnm$ ($Pnma$) phase via intermediate tetragonal $P4/mbm$ one.
}
\end{figure}

The unit cell of the \TFO{} cubic phase ($Pm\overline{3}m$ [\#221, $O^{1}_{h}$], $Z=1$) contains only 5 atoms occupying the Wyckoff positions $1b$ for Tb, $2a$ for Fe, and $3d$ for O as illustrated in Fig.~\ref{fig:structure}(a).
The optimized lattice parameter is equal to $a=3.92$\,\AA.
The group-theoretical analysis of $Pm\overline{3}m$ orthoferrites \RFO{} predicts 5 phonons in the center of the Brillouin zone~\cite{kroumova2003bilbao}
\begin{equation}
\label{eq:group_irrep_total_cubic}
\begin{split} 
\Gamma_{\textrm{total}} = \underbrace{T_{1u}}_{\Gamma_{\textrm{acoustic}}}\!\!\!\oplus \underbrace{3 T_{1u}}_{\Gamma_{\textrm{IR}}} \oplus \underbrace{T_{2u}}_{\Gamma_{\textrm{silent}}},
\end{split}
\end{equation}
The calculation of phonon dispersion reveals a number of imaginary branches, since the cubic phase of \TFO{} is unstable, as shown in Fig.~\ref{fig:DFT_Pm3m_P4mbm}(a).
The lowest imaginary branch is the one with $T_{1u}$ irreducible representation in the center of the Brillouin zone, and
$\mathrm{M}_{4}^{+}$, $\mathrm{R}_{4}^{+}$ and $\mathrm{X}_5^{+}$ at the $\mathrm{M}$, $\mathrm{R}$, $\mathrm{X}$ points of Brillouin zone, respectively [see Fig.~\ref{fig:structure}(d)].
The distortions of the structure by displacements of atoms along eigenvectors of the imaginary modes reduce the crystal space symmetry and yield a number of phases with tetragonal and orthorhombic symmetries.
The most preferable phase $P4/mbm$ was obtained as a phase with the lowest total energy among all possible structures.
The structure was obtained by atomic distortion along the normal coordinate of the phonon with $\mathrm{M}_4^{+}$ irreducible representation at the $\mathrm{M}$ point of the Brillouin zone.
Distortions at the boundary points of the Brillouin zone lead to zone folding, and therefore the volume of the unit cell is increased twice.
The transformation matrix of the structural transition $P_{1}$ is 
\begin{equation}
  \label{eq:trmatrix1}
   P_{1}=
   \left[
   \begin{matrix}
        1 &           -1 & 0  \\
        1 & \phantom{-}1 & 0  \\
        0 & \phantom{-}0 & 1  \\
   \end{matrix}
   \left|
   \,
   \begin{matrix}                                                               
      0  \\                                                                
      0  \\                                                                
      0  \\                                                                
    \end{matrix}                                                                
  \right.                                                                
  \right]                                                                
\end{equation}
where the right column denotes the translation vector.
The correlation diagram for parent-subgroup irreducible representations is plotted in Fig.~\ref{fig:phcorrel}.
It can be clearly seen that the phonon states from the center as well as from the boundary points $\mathrm{X}$, $\mathrm{R}$, $\mathrm{M}$ of the Brillouin zone compose the vibrational states of the tetragonal phase due to Brillouin zone folding. 

The optimized lattice parameters of the tetragonal phase of \TFO{} ($P4/mbm$ [\#127, $D^{5}_{4h}$], $Z=2$) are $a=b=5.49$\,\AA, and $c=3.85$\,\AA, and therefore the phase is more compact with respect to the cubic one as shown in Fig.~\ref{fig:structure}(b).
The tetragonal unit cell contains 10 atoms occupying the Wyckoff positions $2c$ for Tb, $2a$ for Fe, and $2b$ and $4g$ for O.
The group-theoretical analysis of $P4/mbm$ orthoferrites \RFO{} predicts 12 nondegenerate and 9 double-degenerate phonons in the center of the Brillouin zone which decompose by irreducible representations according to the equation~\cite{kroumova2003bilbao}
\begin{equation}
\label{eq:group_irrep_total_tetragonal}
\begin{split} 
\Gamma_{\textrm{total}} = \underbrace{E_{u} \oplus A_{2u}}_{\Gamma_{\textrm{acoustic}}} \oplus \underbrace{A_{g} \oplus B_{1g} \oplus B_{2g} \oplus E_{g}}_{\Gamma_{\textrm{Raman}}}
\oplus \underbrace{7 E_{u} \oplus 3 A_{2u}}_{\Gamma_{\textrm{IR}}} \\
\oplus \underbrace{2 A_{1u} + 2 B_{1u} + A_{2g}}_{\Gamma_{\textrm{silent}}},
\end{split}
\end{equation}
There are several imaginary branches in the calculated phonon dispersion, but this number is reduced as compared to the cubic phase as one might expect; see Fig.~\ref{fig:DFT_Pm3m_P4mbm}(b).
By repeating the stable phase search as described above, the most stable orthorhombic phase $Pbnm$ is established with the transformation matrix as follows:
\begin{equation}
  \label{eq:trmatrix2} 
   P_{2}=
   \left[
   \begin{matrix}
      1 & 0 & 0  \\
      0 & 2 & 0  \\
      0 & 0 & 1  \\
   \end{matrix}
   \left|
   \,
   \begin{matrix}                                                               
      0  \\                                                                
      0  \\                                                                
      \sfrac{1}{2}  \\                                                                
    \end{matrix}                                                                
  \right.                                                                
  \right]
\end{equation}
The tetragonal to orthorhombic phase transition is induced by condensation of the $\mathrm{Z}_{5}^{+}$ phonon at the boundary of the Brillouin zone ($\mathrm{Z}$ point); therefore the structural modification is accompanied by doubling of the unit cell.
According to the correlation diagram in Fig.~\ref{fig:phcorrel} the vibrational states of \TFO{} in the orthorhombic phase are genetically bounded with phonons at the $\Gamma$ and $\mathrm{Z}$ points of the Brillouin zone of the tetragonal phase.

The full sequence of structural transformations from the cubic parent phase to the orthorhombic one includes doubling of the Brillouin zone twice, which leads to a fourfold increase in the number of vibrational modes.
The Brillouin zone folding is usually accompanied by the phonon state mixing, and the correlation diagram in the case of \TFO{} is rather sophisticated due to interaction between phonon branches, which reduces the direct connection between vibrational states in the parent and orthorhombic phases.
The latter explains the multimode influence of TO modes on LO ones, as shown in Fig.~\ref{fig:LOTO_correlation}.

\section{Concluding remarks}
\label{sec:conclusion}

In summary, we have studied the polar optical phonons at the center of the Brillouin zone by the polarized infrared reflectivity technique in the single crystal of rare-earth orthoferrite \TFO{}.
The spectra of the anisotropic complex dielectric permittivity were extracted from the analysis of the experimental results.
The overwhelming majority of predicted TO and LO polar modes were reliably detected according to the polarization selection rules and their parameters were determined. 
To assign the observed TO and LO modes to each other, we supported the experimental study by the DFT calculation of the lattice dynamics.
The obtained frequencies of optical phonons are in fair agreement with the experimental results.
We found that according to the correlation analysis between calculated LO and TO mode eigenvectors most of the LO modes correspond to more than one TO mode at the center of the Brillouin zone due to a strong mode mixing caused by the Coulomb interaction.
However, the Coulomb interaction does not affect the phonons at the Brillouin zone boundaries due to its long-range character.

Next, we have analyzed the dispersion curves of polar phonons throughout the Brillouin zone that were calculated both with and without the Coulomb interaction. 
This allowed us to establish an explicit one-to-one relation between LO and TO polar modes at the center of the Brillouin zone despite the complex correlations of their eigenvectors. 
Furthermore, we found a polar phonon with a negative LO-TO splitting and extended to the Brillouin zone its previously reported general conditions of existence.
Additionally, we completed analysis of lattice dynamics in \TFO{} by measuring angular-resolved polarized Raman scattering from Raman-active phonons.
Using DFT analysis giving a good agreement with the experimental data, we identify which ionic motions contribute to both infrared- and Raman-active phonons.
In particular, we show that $\mathrm{Fe}^{3+}$ ions almost do not contribute to Raman-active phonons, which may account for the apparent absence of spin-phonon effects on the frequencies of these phonons reported in the literature.
We believe that our results will stimulate further research into nonlinear phononic and magnetophononic effects in the rare-earth orthoferrites \RFO{} since they explicitly show a degree of mixing between different phonon modes~\cite{nova2017effective,juraschek2017ultrafast,afanasiev2021ultrafast,huang2024extreme,zhang2024upconversion,zhang2024coupling}.

\section*{Acknowledgments}
The single crystals used in the experiments were grown by A.\,M.\,Balbashov.
We thank M.\,P.\,Scheglov and N.\,A.\,Arkhipov for the help with the x-ray orientation of single crystals.
This work was supported by the Russian Science Foundation under Grant no.\,22-72-00025, https://rscf.ru/en/project/22-72-00025/.
A.I.B. acknowledges the support of the Ministry of Science and Higher Education of the Russian Federation (Grant No. FSWR-2024-0003).
N.N.N. and K.N.B. acknowledge support by Research Project No. FFUU-2022-0003 of the Institute of Spectroscopy of the Russian Academy of Sciences.
V.A.C. acknowledges support by the Ministry of Science and Higher Education of the Russian Federation, Project No.~FEUZ-2023-0017.
R.V.M. acknowledges the support of the Royal Society International Exchanges 2021, Grant No. IES$\backslash$R2$\backslash$212182.

\bibliography{bibliography}

\end{document}


\title{{S}upplemental {M}aterial to:\\Lattice dynamics and mixing of polar phonons in the rare-earth orthoferrite TbFeO$_{3}$}

\author{R.~M.~Dubrovin\,\orcidlink{0000-0002-7235-7805}}
\email{dubrovin@mail.ioffe.ru}
\affiliation{Ioffe Institute, Russian Academy of Sciences, 194021 St.\,Petersburg, Russia}
\author{E.~M.~Roginskii\,\orcidlink{0000-0002-5627-5877}}
\affiliation{Ioffe Institute, Russian Academy of Sciences, 194021 St.\,Petersburg, Russia}
\author{V.~A.~Chernyshev\,\orcidlink{0000-0002-3106-3069}}
\affiliation{Department of Basic and Applied Physics, Ural Federal University, 620002 Yekaterinburg, Russia}
\author{N.~N.~Novikova\,\orcidlink{0000-0003-2428-6114}}
\affiliation{Institute of Spectroscopy, Russian Academy of Sciences, 108840 Moscow, Troitsk, Russia}
\author{M.~A.~Elistratova\,\orcidlink{0000-0002-1573-1151}}
\affiliation{Ioffe Institute, Russian Academy of Sciences, 194021 St.\,Petersburg, Russia}
\author{I.~A.~Eliseyev\,\orcidlink{0000-0001-9980-6191}}
\affiliation{Ioffe Institute, Russian Academy of Sciences, 194021 St.\,Petersburg, Russia}
\author{A.~N.~Smirnov\,\orcidlink{0000-0001-9709-5138}}
\affiliation{Ioffe Institute, Russian Academy of Sciences, 194021 St.\,Petersburg, Russia}
\author{A.~I.~Brulev\,\orcidlink{0009-0004-5339-8486}}
\affiliation{Ioffe Institute, Russian Academy of Sciences, 194021 St.\,Petersburg, Russia}
\affiliation{University of Nizhny Novgorod, 603022 Nizhny Novgorod, Russia}
\author{K.~N.~Boldyrev\,\orcidlink{0000-0002-3784-7294}}
\affiliation{Institute of Spectroscopy, Russian Academy of Sciences, 108840 Moscow, Troitsk, Russia}
\author{V.~{Yu}.~Davydov\,\orcidlink{0000-0002-5255-9530}}
\affiliation{Ioffe Institute, Russian Academy of Sciences, 194021 St.\,Petersburg, Russia}
\author{R.~V.~Mikhaylovskiy\,\orcidlink{0000-0003-3780-0872}}
\affiliation{Department of Physics, Lancaster University, Bailrigg, Lancaster LA1 4YW, United Kingdom}
\author{A.~M.~Kalashnikova\,\orcidlink{0000-0001-5635-6186}}
\affiliation{Ioffe Institute, Russian Academy of Sciences, 194021 St.\,Petersburg, Russia}
\author{R.~V.~Pisarev\,\orcidlink{0000-0002-2008-9335}}
\affiliation{Ioffe Institute, Russian Academy of Sciences, 194021 St.\,Petersburg, Russia}

\date{\today}

\maketitle

\onecolumngrid

\section*{Methods}
\label{sec:methods}

The single crystal of orthoferrite \TFO{} was grown by the floating zone technique with a light heating as described in detail in Refs.~\cite{balbashov2019contemporary,balbashov1981apparatus}. 
The single crystals were oriented by the x-ray diffraction in the reflection geometry using monochromatic Cu K$\alpha_{1}$ (1.5406\,\AA) radiation.
The oriented single crystals were cut into samples with normal of the surface along the three main crystallographic axes and polished to the optical surface quality.
The samples have a typical thickness of about 700\,$\mu$m and a surface size of about $7\times{7}$\,mm$^{2}$.

The infrared reflectivity measurements were carried out at room temperature with near normal incident light (the incident light beam was at 10$^{\circ}$ from the normal to the sample surface) using a Bruker IFS 66v/S spectrometer with DTGS (50--450\,cm$^{-1}$) and DLaTGS (450--5000\,cm$^{-1}$) detectors with a resolution of 4\,cm$^{-1}$, a number of scans of 128 and a scanning velocity of 2.2\,kHz.
The linear polarization of light emitted by the globar source was set by the THz linear thin film polarizer along the main crystallographic axes of the crystal samples.
Absolute values of reflectivity were obtained by normalizing the spectra obtained from the samples by the reference one from the gold mirror.  

The polarized Raman spectra were measured in the range from 90 to 1000\,cm$^{-1}$ with a resolution of $<$1\,cm$^{-1}$ using the Horiba LabRAM HREvo UV-Vis-NIR-Open spectrometer equipped with a liquid nitrogen cooled CCD detector.
For excitation, a 532\,nm line of a Nd:YAG laser Torus (Laser Quantum) was used with low power of 4\,mW thus avoiding overheating of optically dense samples. 
The experiments were performed at ambient conditions in the backscattering geometry using an Olympus MPLN 100$\times$ objective employed both to focus the excitation beam into a spot with a diameter of $<1$ $\mu$m and to collect the scattered light.
Polarization of incident and scattered light was controlled using a half-wave plate installed in a motorized rotator.
The polarized light components were discriminated by a film analyzer installed in the path of the scattered light.

The experimental results were supported by lattice dynamics calculations of \TFO{} from the first principles. 
We have applied the density functional theory (DFT) within the Perdew-Burke-Ernzerhof (PBE) parameterized generalized gradient approximation (GGA)~\cite{perdew1996generalized} as implemented in the Vienna \textit{ab initio} simulation package (\textsc{VASP})~\cite{kresse1996efficiency,kresse1996efficient}. 
The plane-wave basis projector augmented wave (PAW) pseudopotentials with valence electronic states $\mathrm{Tb}$ ($5p^{6}5d^{1}6s^{2}$) with the $4f$ electrons frozen in the ionic core, $\mathrm{Fe}$ ($3d^{7}4s^{1}$), and $\mathrm{O}$ ($2s^{2}2p^{4}$) were used.
The $3d$ states of $\mathrm{Fe}$ were corrected through the DFT + $U$ ($U=4$\,eV) approximation within the Dudarev formalism~\cite{dudarev1998electron}. 
The calculations were converged with a ${8}\times{6}\times{8}$ $k$-points mesh for sampling of the reciprocal space from the Monkhorst-Pack scheme~\cite{monkhorst1976special} and a plane wave energy cut-off of 650\,eV.
The phonon dispersion curves over the entire Brillouin zone in the harmonic approximation were obtained by calculating forces in the $2\times2\times2$ supercell with finite atomic displacements technique implemented in the \textsc{Phonopy} package~\cite{togo2015first}. 
To reveal the LO-TO splitting due to the long-range Coulomb interaction the nonanalytic correction to the dynamical matrix was applied in the Wang approach~\cite{wang2010mixed}.
The Raman spectra were simulated using Raman tensor components calculated within Coupled Perturbed Hartree-Fock (CPHF/KS) approach~\cite{maschio2012infrared,maschio2013raman} as implemented in \textsc{CRYSTAL14} package~\cite{dovesi2014crystal14}.
These calculations were performed with B3LYP hybrid functional~\cite{becke1993density} using the quasi-relativistic ECP$n$MWB pseudo-potentials for $\mathrm{Tb}^{3+}$ ($n=54$) to describe the core electrons including $4f$ shell~\cite{dolg1989energy,dolg1993combination}.
To describe the valence electrons of rare-earth ions the ECP$n$MWB-I basis sets were used~\cite{dolg1989energy,yang2005valence}. 
All electron basis sets have been used to describe ions with a contraction scheme of (8s)-(6411sp)-(4111d)-(1f) for $\mathrm{Fe}$, and (8s)-(51sp)-(1d) for $\mathrm{O}$~\cite{peintinger2013consistent}.
The ferromagnetic configuration of $\mathrm{Tb}$ and $\mathrm{Fe}$ spins was used in the calculations.



\section*{Infrared spectroscopy}

The reflectivity spectra were analyzed using the Kramers--Kronig constrained variational technique implemented in the \textsc{RefFIT} software~\cite{kuzmenko2005kramers} which allowed us to obtain spectra of the complex dielectric permittivity $\varepsilon(\omega) = \varepsilon_{1}(\omega) - i\varepsilon_{2}(\omega)$. 
Peaks in the spectra of imaginary part of the dielectric permittivity $\Im[\varepsilon(\omega)]$ and the inverse dielectric permittivity $\Im[\varepsilon^{-1}(\omega)]$ correspond to the frequencies of transverse (TO) and longitudinal (LO) polar phonons, respectively~\cite{schubert2004infrared}.
Further, using these frequencies, the reflectivity spectra for all polarizations were fitted by using the factorized form of the complex dielectric permittivity~\cite{gervais1974anharmonicity}
\begin{equation}
\label{eq:epsilon_TOLO}
\varepsilon(\omega) = \varepsilon_{1}(\omega) - i\varepsilon_{2}(\omega) = \varepsilon_{\infty}\prod\limits_{j}\frac{{\omega^{2}_{j\textrm{LO}}} - {\omega}^2 + i\gamma_{j\textrm{LO}}\omega}{{\omega^{2}_{j\textrm{TO}}} - {\omega}^2 + i\gamma_{j\textrm{TO}}\omega},
\end{equation}
where $\varepsilon_{\infty}$ is the high-frequency dielectric permittivity, $\omega_{j\textrm{TO}}$, $\omega_{j\textrm{LO}}$, $\gamma_{j\textrm{TO}}$ and $\gamma_{j\textrm{LO}}$ correspond to $\textrm{TO}$ and $\textrm{LO}$ frequencies ($\omega_{j}$) and dampings ($\gamma_{j}$) of the $j$th polar phonon of the specific symmetry, respectively.
Multiplication occurs over all polar phonons with the specific symmetry which are active for this polarization of the incident light.
Eq.~\eqref{eq:epsilon_TOLO} at the $\omega=0$ converges to the well-known Lyddane-Sachs-Teller relation~\cite{lyddane1941polar}. 
For normal incidence, the infrared reflectivity $R(\omega)$ and complex dielectric function $\varepsilon(\omega)$ are related to each other via the Fresnel equation~\cite{born2013principles}
\begin{equation}
\label{eq:reflectivity}
R(\omega) = \Bigl|\frac{\sqrt{\varepsilon(\omega)} - 1}{\sqrt{\varepsilon(\omega)} + 1}\Bigr|^2.
\end{equation}

\section*{Raman spectroscopy}
\label{sec:raman}

The scattering tensors of Raman-active phonons for orthoferrites in the $Pbnm$ setting of crystal axes are defined as~\cite{martin2001melting,kroumova2003bilbao} 
\begin{equation}
    \label{eq:raman_tensors}
    \begin{gathered}
        \mathcal{R}_{A_{g}} =
        \begin{pmatrix}
             a& 0 & 0 \\ 
             0& b & 0 \\ 
             0& 0 & c
        \end{pmatrix}
        ,\quad
        \mathcal{R}_{B_{1g}} = 
        \begin{pmatrix}
             0& d & 0 \\ 
             d& 0 & 0 \\
             0& 0 & 0
        \end{pmatrix},\\[1ex]
        \mathcal{R}_{B_{2g}} = 
        \begin{pmatrix}
             0& 0 & e \\ 
             0& 0 & 0 \\
             e& 0 & 0
        \end{pmatrix}
        ,\quad
        \mathcal{R}_{B_{3g}} = 
        \begin{pmatrix}
             0& 0 & 0 \\ 
             0& 0 & f \\
             0& f & 0
        \end{pmatrix},
    \end{gathered}
\end{equation}
where $a$--$f$ are independent tensor elements.
The intensity of the light scattered by a Raman-active phonon is determined from the Raman tensor by the following relation~\cite{loudon2001raman}
\begin{equation}
\label{eq:raman}
I \propto |\bm{e}_{s} \mathcal{R} \bm{e}_{i}|^2,
\end{equation}
where $\bm{e}_{s}$ and $\bm{e}_{i}$ are the unit vectors of the scattered and incident light polarizations, respectively. 
Thus, Raman-active phonons of a given symmetry can be selectively distinguished by using specific polarization configurations of incident and scattered light with respect to the main crystallographic axes of the crystal.
The polarization configuration is usually given in Porto's notation, according to which a set of four symbols is used, $\bm{k}_{i} (\bm{E}_{i}\bm{E}_{s}) \bm{k}_{s}$, where $\bm{k}_{i}$ and $\bm{k}_{s}$ are the directions of the propagation of the incident and scattered light (for the backscattering geometry these directions are opposite $\bm{k}_{i}=\overline{\bm{k}}_{s}$), while $\bm{E}_{i}$ and $\bm{E}_{s}$ are the polarization of the incident and scattered light in the crystal axes coordinate system~\cite{damen1966raman}.
According to the Eqs.~\eqref{eq:raman_tensors} and~\eqref{eq:raman}, the $A_{g}$ phonons are active for the parallel polarization settings $\bm{e}_{i}\parallel{\bm{e}_{s}}$ 
whereas $B_{1g}$, $B_{2g}$ and $B_{3g}$ phonons are distinguishable in the crossed configurations $\bm{e}_{i}\perp{\bm{e}_{s}}$.

\begin{figure*}
\centering
\includegraphics[width=1\columnwidth]{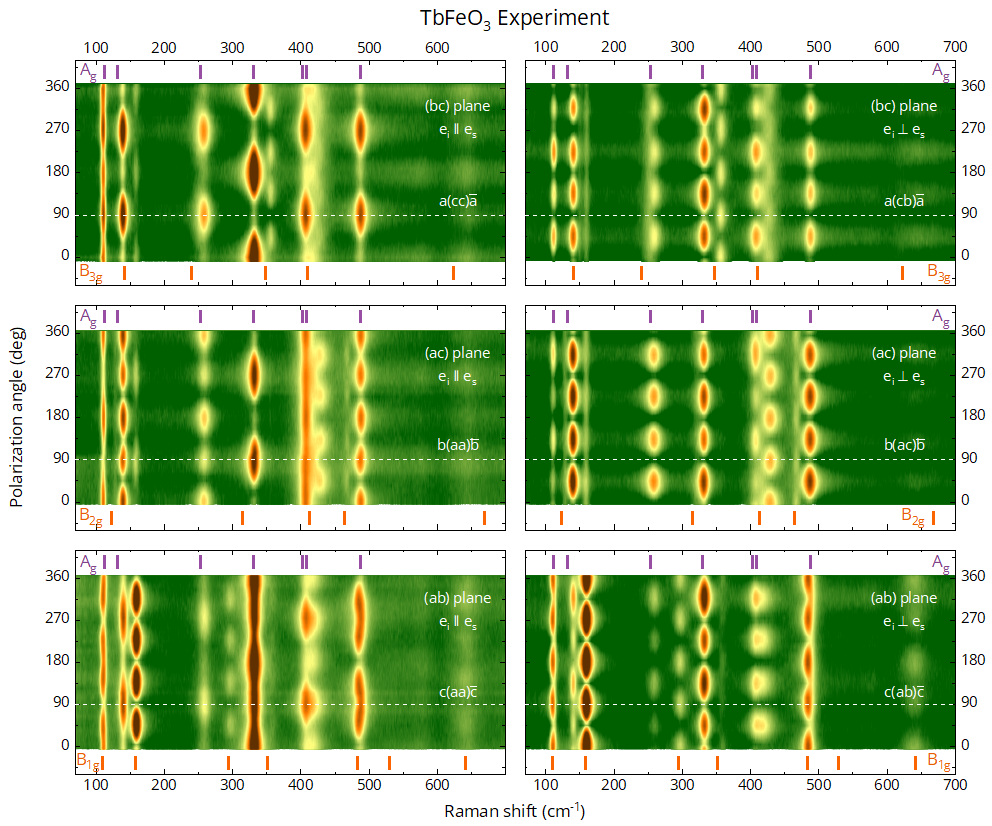}
\caption{\label{fig:raman_angle_experiment_map}
Experimental Raman scattering log-scale intensity color maps as a function of Raman shift and polarization angle between the incident light $\bm{e}_{i}$ and the selected main crystallographic axis in parallel  configurations.
The dashed white lines correspond to the specified polarization configuration in Porto's notation.
Color sticks in the each plot present the calculated phonon frequencies.
}
\end{figure*}

\begin{figure*}
\centering
\includegraphics[width=1\columnwidth]{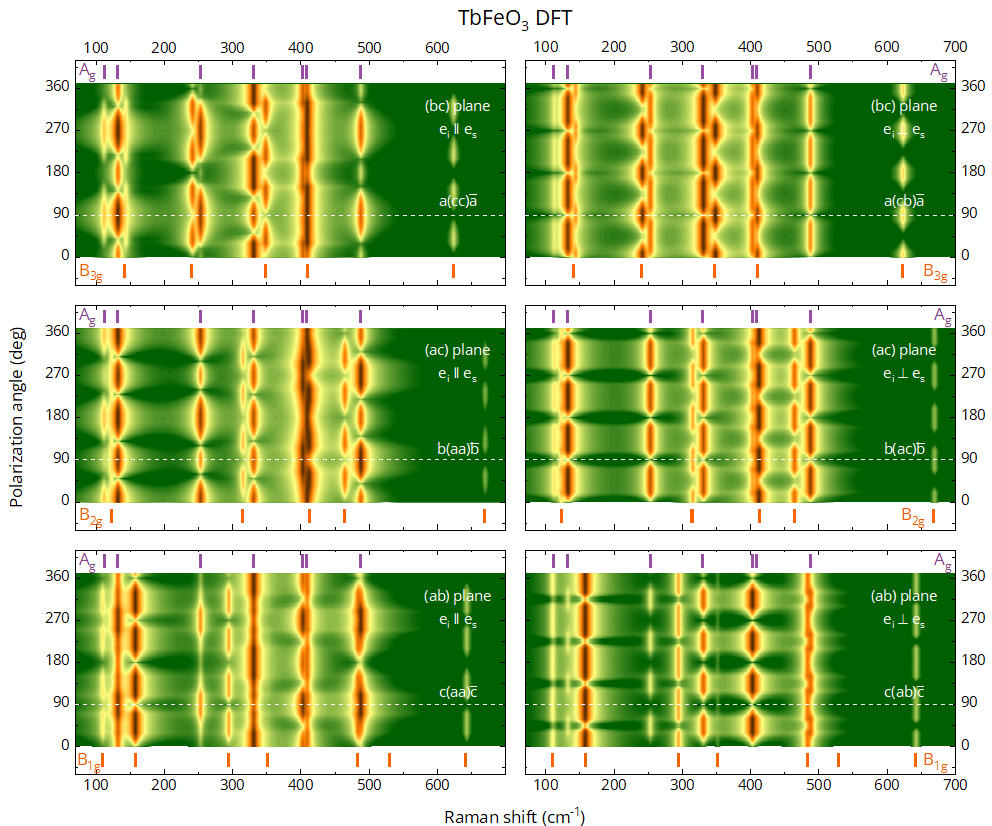}
\caption{\label{fig:raman_angle_dft_map}
Calculated Raman scattering log-scale intensity color maps as a function of Raman shift and polarization angle between the incident light $\bm{e}_{i}$ and the selected main crystallographic axis in parallel ($\bm{e}_{i}\parallel{\bm{e}_{s}}$) and crossed ($\bm{e}_{i}\perp{\bm{e}_{s}}$) configurations.
The dashed white lines correspond to the specified polarization configuration in Porto's notation.
Color sticks in the each plot present the calculated phonon frequencies.
}
\end{figure*}

\begin{figure*}
\centering
\includegraphics[width=1\columnwidth]{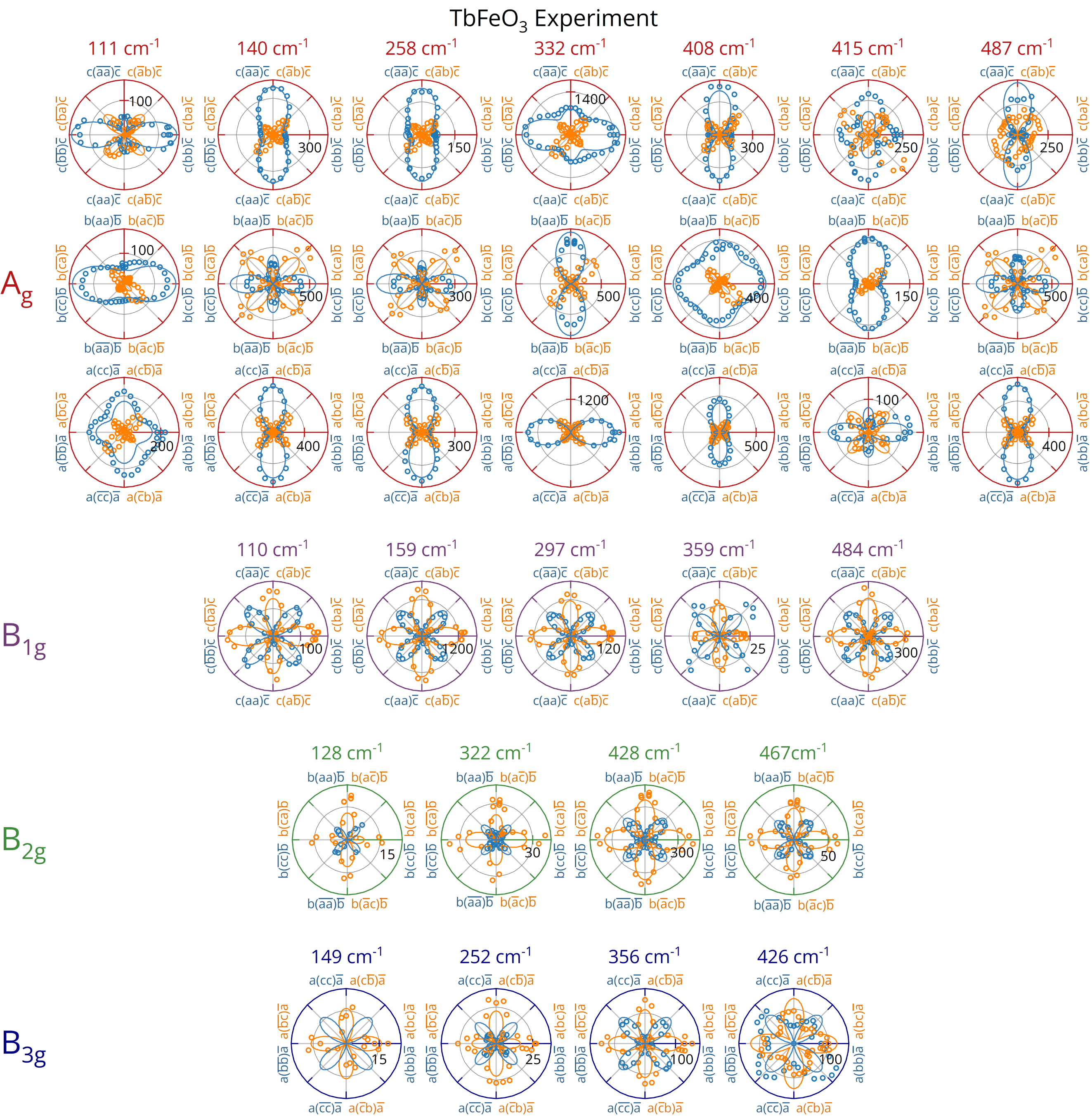}
\caption{\label{fig:raman_angle_experiment}
Experimental angle dependences of the intensities of the Raman-active $A_{g}$, $B_{1g}$, $B_{2g}$, and $B_{3g}$ phonons in parallel ($\bm{e}_{i}\parallel{\bm{e}_{s}}$) and crossed ($\bm{e}_{i}\perp{\bm{e}_{s}}$) polarization configurations in \TFO. 
}
\end{figure*}

\begin{figure*}
\centering
\includegraphics[width=1\columnwidth]{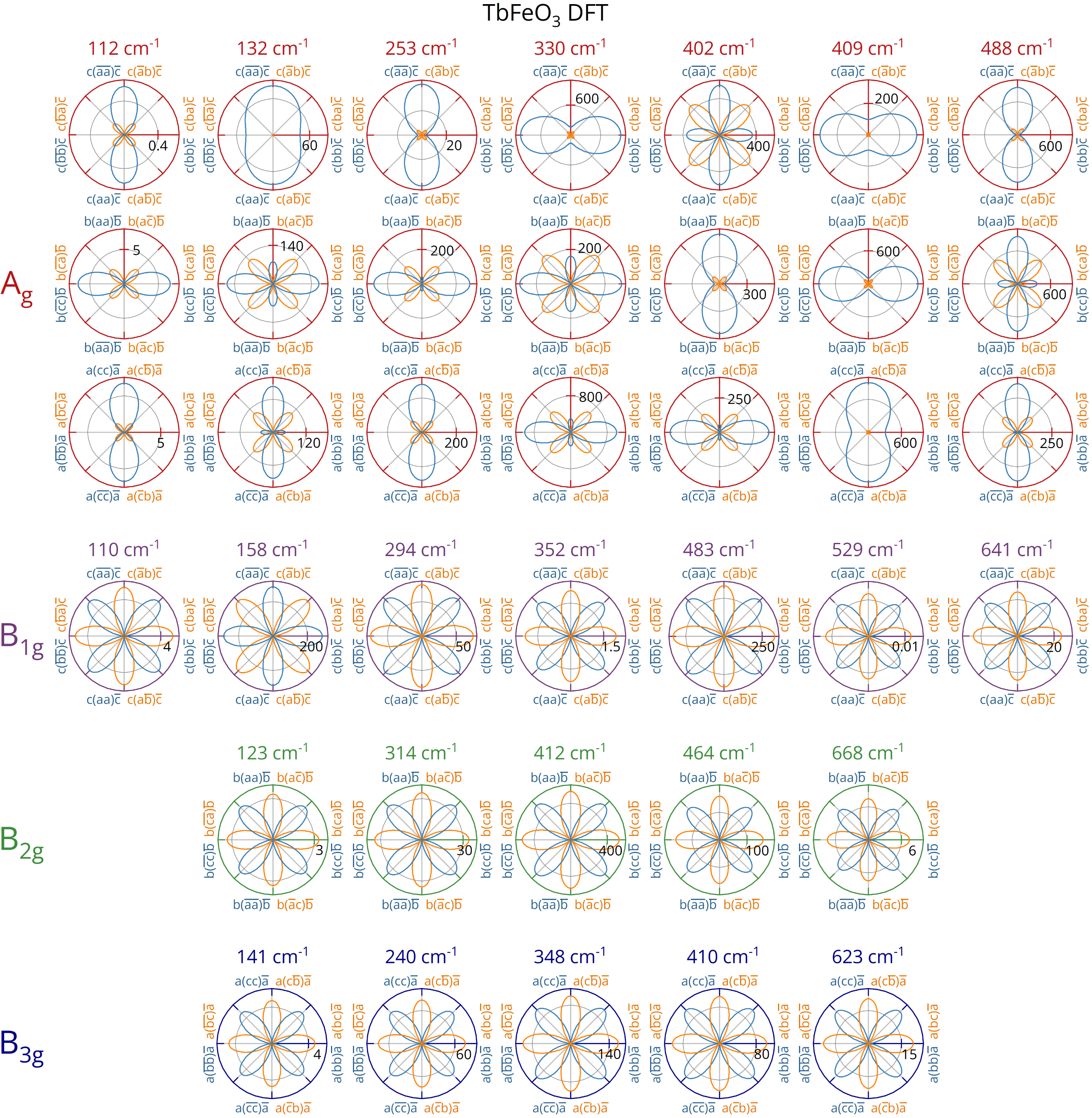}
\caption{\label{fig:raman_angle_dft}
Calculated angle dependences of the intensities of the Raman-active $A_{g}$, $B_{1g}$, $B_{2g}$, and $B_{3g}$ phonons in parallel ($\bm{e}_{i}\parallel{\bm{e}_{s}}$) and crossed ($\bm{e}_{i}\perp{\bm{e}_{s}}$) polarization configurations in \TFO.
}
\end{figure*}

\bibliography{bibliography}